\documentclass[12pt,preprint]{aastex}

\slugcomment{Based on observations carried out with the IRAM Plateau de Bure Interferometer.
IRAM is supported by INSU/CNRS (France), MPG (Germany) and IGN (Spain).}

\shorttitle{AB Aur mm-observations and modeling}
\shortauthors{Semenov et al.}

\begin{document}

\title{Millimeter observations and modeling of the AB Aurigae system}

\author{D. Semenov}
\affil{Max Planck Institute for Astronomy,
K\"onigstuhl 17, D-69117 Heidelberg, Germany}
\email{semenov@mpia.de}

\author{Ya. Pavlyuchenkov}
\affil{Institute of Astronomy of the Russian Academy of Sciences,
48, Pyatnitskaya Str., Moscow, 109017, Russia}
\email{pavyar@inasan.rssi.ru}

\author{K. Schreyer}
\affil{Astrophysical Institute and University Observatory,
Schillerg\"a{\ss}chen 2-3, D-07745 Jena, Germany}
\email{martin@astro.uni-jena.de}

\author{Th. Henning}
\affil{Max Planck Institute for Astronomy,
K\"onigstuhl 17, D-69117 Heidelberg, Germany}
\email{henning@mpia.de}

\author{K. Dullemond}
\affil{Max Planck Institute for Astronomy,
K\"onigstuhl 17, D-69117 Heidelberg, Germany}
\email{dullemon@mpia.de}

\and

\author{A. Bacmann}
\affil{Observatoire de Bordeaux,
2 rue de l'Observatoire, BP 89 F-33270 Floirac, France}
\email{bacmann@obs.u-bordeaux1.fr}

\begin{abstract}
We present the results of millimeter observations and a suitable chemical and radiative transfer
model of the \objectname{AB Aurigae} (\objectname{HD 31293}) circumstellar disk and surrounding
envelope. The integral molecular content of this system is studied by observing CO, C$^{18}$O, CS,
HCO$^+$, DCO$^+$, H$_2$CO, HCN, HNC, and SiO rotational lines with the IRAM 30-m antenna, while the
disk is mapped in the HCO$^+$(1-0) transition with the Plateau de Bure interferometer. Using a
flared disk model with a vertical temperature gradient and an isothermal spherical envelope model
with a shadowed midplane and two unshielded cones together with a gas-grain chemical network,
time-dependent abundances of observationally important molecules are calculated. Then a 2D
non-LTE line radiative transfer code is applied to compute excitation temperatures of several
rotational transitions of HCO$^+$, CO, C$^{18}$O, and CS molecules. We synthesize the
HCO$^+$(1-0) interferometric map along with single-dish CO(2-1), C$^{18}$O(2-1), HCO$^+$(1-0),
HCO$^+$(3-2), CS(2-1), and CS(5-4) spectra and compared them with the observations. Our disk model
successfully reproduces observed interferometric HCO$^+$(1-0) data, thereby constraining
the following disk properties: (1) the inclination angle $\iota=17^{+6}_{-3}\degr$, (2) the
position angle $\phi=80\pm30\degr$, (3) the size $R_\mathrm{out}=400\pm200$~AU, (4) the mass 
$M_\mathrm{disk}=1.3\cdot10^{-2}\,M_{\sun}$ (with a factor of $\sim7$ uncertainty), and (5) that the 
disk is in Keplerian rotation. Furthermore, indirect evidence for a local inhomogeneity of the
envelope at $\ga600$~AU is found. The single-dish spectra are synthesized for three
different cases, namely, for the disk model, for the envelope model, and for their combination.
An overall reasonable agreement between all modeled and acquired line intensities, widths, and
profiles is achieved for the latter model, with the exception of the CS(5-4) data that require
presence of high density clumpy structures in the model. This allows 
to constrain the physical structure of the AB Aur inner envelope: 
(1) its mass-average temperature is about $35\pm14$~K, (2) the density goes inversely down with the radius, 
$\rho \propto r^{-1.0\pm0.3}$, starting from an initial value  $n_0 \approx 3.9\cdot10^5$~cm$^{-3}$ at 
$400$~AU, and (3) the mass of the shielded region within 2\,200~AU is about $4\cdot10^{-3}\,M_{\sun}$ 
(the latter two quantities are uncertain by a factor of $\sim7$). Also, evolutionary nature and lifetime for 
dispersal of the AB Aur system and Herbig Ae/Be systems in general are discussed.
\end{abstract}


\section{Introduction}
Nowadays the study of planet formation attracts particular attention in astrophysics since the
recent discovery of extrasolar planets \citep[e.g.,][]{MB00}. One of the main goals of this study
is to envisage the evolutionary scenario for protoplanetary disks around young
low- and intermediate-mass stars in which planets might have formed or are forming \citep[e.g.,][]{BHN00}.
High-resolution observations with infrared and millimeter interferometers as well as UV-to-cm
ground-based and space-borne single-dish telescopes have clearly demonstrated that such disks exist and provided 
a wealth of information concerning their gas and dust content
\citep[see, e.g.,][]{TK91,Kawaea93,ME94,DGG97,vDB98,Pea99,vdAea00,Tea01,Wea02,Wahea03,Aea04}. On the other
hand, increasing computer power and the development of sophisticated numerical algorithms during the last
years have made possible a more realistic modeling of their hydrodynamical and chemical evolution
\citep[e.g.,][]{Gail98,WKMH98,Aea02,Mea02,vZea03,IHMM04,Red2}. Coupled with appropriate (line) radiative
transfer simulations, this offers an unique way to account for relevant observational data not only
qualitatively but also quantitatively \citep[e.g.,][]{GD98,HvT00,Hog01,vZea01,BCDH03}.

The aim of the present paper is to develop the first comprehensive model of the physical, chemical, and dynamical 
structure of the circumstellar matter orbiting around an intermediate-mass Herbig Ae star based on available
observations. We focus on AB Aur, which is one of the nearest and best-studied object among the entire
class of the Herbig Ae stars. The AB Aur system consists of a pre-main-sequence (PMS) star with spectral type
A0Ve+sh \citep[mass $M_*\sim 2.4M_\sun$, radius $R_*\approx 2.5R_\sun$, ][]{Tea94} located at a distance of
$\approx 145$~pc \citep{vdAea97,vdAea98}, which is surrounded by a $r\la 450$~AU rotating
circumstellar disk \citep{MS97} and an extended ($r>1\,000$~AU) envelope \citep{NG95,Grea99,Rea01}.
As it is a PMS object, the age of AB Aur can be roughly estimated by using evolutionary track modeling, 
giving a value of $t\approx2$-5~Myr \citep[][]{Tea01,dWea03}.

One of the long-standing mysteries that is related to this object is the value of the disk inclination angle.
\citet{MS97} have spatially resolved a disk-like configuration around AB Aur using OVRO ({\it Owens
Valley Radio Observatory}) aperture-synthesis imaging in the $^{13}$CO(1-0) line. From the aspect ratio
of the disk major and marginally resolved minor semiaxes, $\sim 110$~AU/450~AU, they have derived a
disk inclination angle $\iota\approx76\degr$ (almost ``edge-on'' orientation), which is in close agreement 
with mid-infrared (MIR) observations of \citet{Mea95}. However, near-infrared (NIR) interferometric 
observations by \citet{MGea99,MGea01} have revealed that the AB Aur system looks nearly spherically-symmetric 
at $\sim1$~AU scale, implying that the inclination angle is close to $0\degr$ (``face-on''). This is in accordance with a
value of $\iota<45\degr$ deduced from scattered-light {\it Hubble Space Telescope} (HST) coronographic imaging by
\citet{Grea99} and $\iota\in[27\degr,35\degr]$ estimated by \citet{Eea03} using NIR measurements 
with the {\it Palomar Testbed Interferometer} (PTI). Recently, using coronographic NIR-imaging with the {\it Subaru} 
telescope, \citet{Fea04} have found an inclination $\iota=30\pm5\degr$. Moreover, \citet{Mea99} have 
obtained $\iota=80\degr$ in the framework 
of their best-fit ``disk-in-envelope'' model to the observed spectral energy distribution (SED). A similar value 
of $65\degr$ has been used by \citet{Dea03}, who have successfully fitted the AB Aur SED by a flared 
disk model with a puffed-up inner rim. Obviously, there is a general disagreement regarding the disk orientation
observed with IR- and mm-interferometers or derived from the SED modeling, though more recent studies point 
to a face-on configuration. A combination of high-quality observational data and appropriate
theoretical modeling would allow to determine precisely, among many other parameters, the orientation of the AB Aur 
system.

In this paper, we report on the results of our observations at millimeter wavelengths and modeling of the 
AB Aur system. The object was observed at low resolution ($\sim 10\arcsec$-$30\arcsec$) using the IRAM 30-m 
antenna between 2000 and 2001 and at a higher $\approx 5\arcsec$ resolution in the HCO$^+$(1-0) line with 
the IRAM {\it Plateau de Bure Interferometer} (PdBI) in 2002. About a dozen rotational transitions of CO, C$^{18}$O, CS, 
HCN, HNC, HCO$^+$, DCO$^+$, SiO, and H$_2$CO were detected with the IRAM 30-m telescope. These single-dish and 
interferometric data as well as supplementary data from the literature form the observational basis for our
study.

First, we simulate the time-dependent chemical evolution of the AB Aur system using a gas-grain chemical network and 
assuming a two-component model of a flared passive accretion disk enshrouded in a diffuse spherical
envelope. Second, we apply a 2D non-LTE line radiative transfer code to translate the calculated
molecular abundances of CS, CO, and HCO$^+$ to the corresponding synthetic beam-convolved single-dish
and interferometric spectra. Next, we compare the observed and synthesized emission lines in a systematic way 
in order to determine the model parameters consequently, each after another. Iterating these three stages of the modeling, 
we find the best-fit model of the AB Aur disk and envelope and estimate uncertainties of the constrained parameters. 

Our primary intention is to verify the strength of such an advanced theoretical approach to account for various observed 
interferometric and single-dish molecular spectra {\em simultaneously}. As a by-product of this study, many important 
parameters describing the physical, chemical, and dynamical structure of the AB Aur system can be constrained 
{\em independently} from other investigations performed so far.

This paper is organized as follows. In Sect.~\ref{obs} we describe our single-dish and interferometric
millimeter observations of AB Aur with the PdBI array and IRAM 30-m antenna. The relevant physical
and chemical models of the disk and envelope are presented and discussed in Sect.~\ref{mod}. We briefly
outline the algorithm and limitations of the 2D line radiative transfer code ``URAN(IA)'' and calculate
excitation temperatures of several rotational transitions of the CS, CO, C$^{18}$O, and HCO$^+$ molecules in
Sect.~\ref{lrt}. Using the best-fit model of the AB Aurigae system, the corresponding interferometric map of the 
HCO$^+$(1-0) emission and single-dish CO(2-1), HCO$^+$(1-0), HCO$^+$(3-2), C$^{18}$O(2-1), and CS(2-1) spectra are 
synthesized and compared with the observed spectra in Sect.~\ref{res}. In this Section, we also discuss how parameters
of the best-fit model and their uncertainties are constrained, and what the evolutionary status and lifetime of the 
AB Aur system are. A summary and final conclusions follow in Sect.~\ref{con}.

\section{Millimeter observations of AB Aur}
\label{obs}

\subsection{The IRAM 30-m data}
\label{iram}
The observations of AB Aur with the IRAM 30-m dish were performed during two runs in 
September 2000 and October 2001. We measured several spectral
line settings at the following sky position:\\
$\alpha_{2000}$ = 04$^{\rm h}$ 55$^{\rm m}$ 45$^{\rm s}$.8,\\
$\delta_{2000}$ = $+30\degr$ 33$\arcmin$ 04$\arcsec$.3.

The observations were carried out with all four receivers, A, B, C, D as the
frontends and the autocorrelator as the backend. We used the autocorrelator split into different
subbands with resolution between 20~kHz and 80~kHz, depending on the presence of (hyper-) fine
structure in the transition under investigation. During the first run, the data were obtained applying both the beam
wobbling mode with a beamthrow of 240$\arcsec$ and the frequency switching mode with a
frequency shift of 5~MHz. It turned out that at the position 240$\arcsec$ away from the center the
emission from the nearby remnant nebulosity is still strong enough to contaminate the observations. 
Therefore, during the second set of observations, the same frequency switching
and a position switching with a larger $20\arcmin$ offset position were used. The
typical spectral resolution was about 0.2~km\,s$^{-1}$. The total integration time varied between
5~min for CO and $\sim45$~min for HCN, depending on the signal strength. The pointing of
the instrument was checked about every $1.5$ hour, which resulted in a typical error of
$\la 5\arcsec$. In the case of several measurements of a particular molecular transition with the
on-off and frequency switching techniques, we relied always on that which had the
highest signal-to-noise ratio. The chopper-wheel method was applied to
calibrate the spectra in units of the antenna temperature $T^*_{\rm A}$. To reduce all
spectroscopic data, the standard GILDAS software package was used.

We summarize all single-dish detections in Table~\ref{lines}. In total, nine different molecular species
were detected: CO, C$^{18}$O, CS, HCO$^+$, DCO$^+$, H$_2$CO, HCN, HNC, and SiO. The
observed intensities (mJy) were converted to the antenna temperatures (Kelvin) and afterwards to the main beam temperatures:
$T_{\rm mb}$ = $T^*_{\rm A}/\eta_{\rm mb}$. The beam efficiency values, $\eta_{\rm mb}$, were
taken from the IRAM Newsletter, N.~18, 1994 (see also Table~\ref{lines}, Col.~10). In addition,
the following spectral lines were only marginally detected (with $\la 2$ sigma):
CN($J$=3/2-1/2,~$F$=5/2-3/2), H$_2$CO($J_{\rm K_p,K_o}$=$2_{1,2}$-$1_{1,1}$), and
H$_2$CO($J_{\rm K_p,K_o}$=$3_{0,3}$-$2_{0,2}$). Other observationally interesting lines,
like HC$_3$N($J$=10-9), N$_2$H$^+$($J$=1-0), C$_2$H($J$=3/2-1/2),
CH$_3$OH($J_{\rm K_p,K_o}$=$2_{1,1}$-$1_{1,0}$), and CH$_3$CN($J$=5-4,~$K$=0) were not detected at all.

\begin{figure}
\includegraphics[width=0.95\textwidth,clip=]{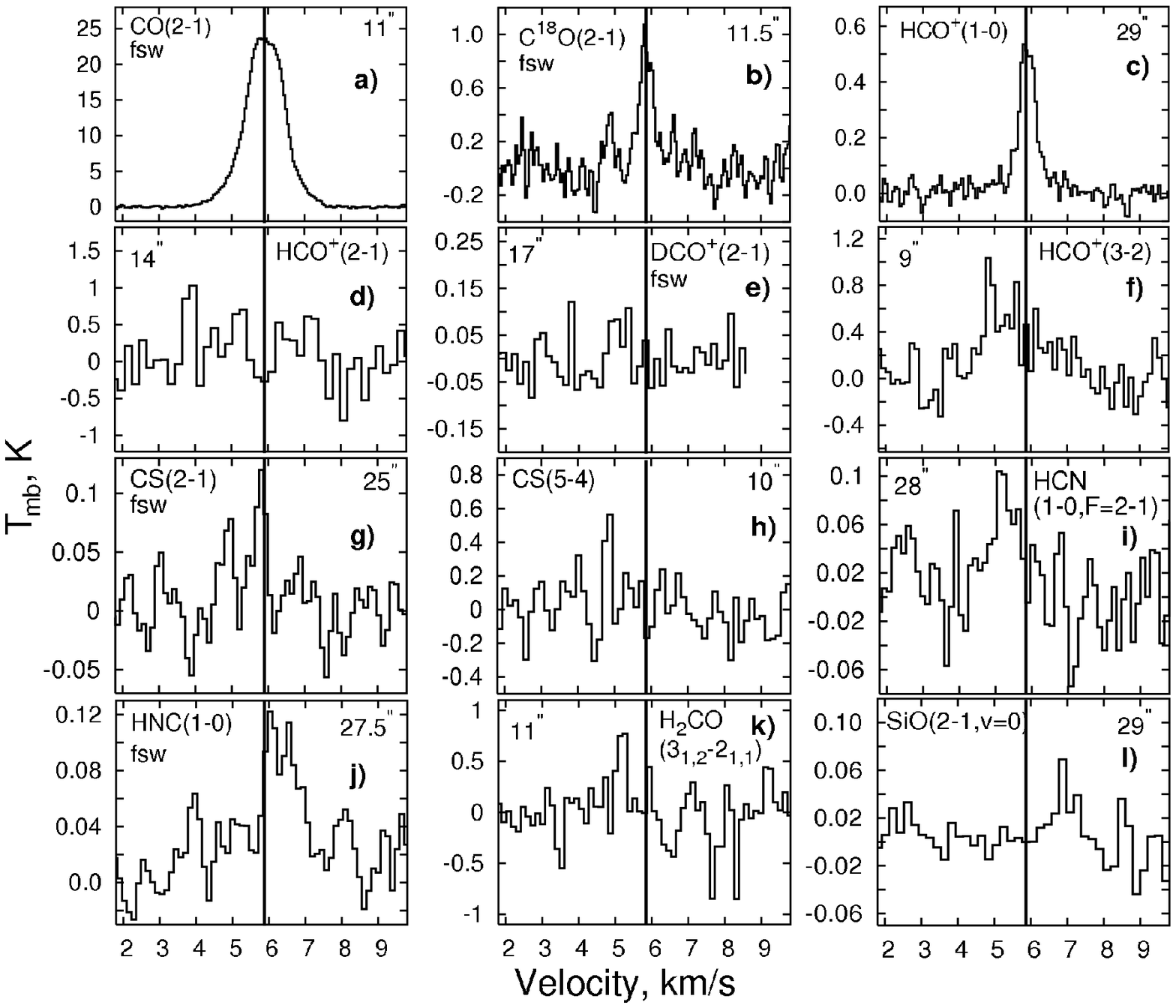}
\caption{Single-dish emission lines observed toward AB Aur
with the IRAM 30-m antenna. From the upper left to the lower right panel ({\bf a-l}) the following
lines are shown: CO(2-1), C$^{18}$O(2-1), HCO$^+$(1-0), HCO$^+$(2-1), DCO$^+$(2-1),
HCO$^+$(3-2), CS(2-1), CS(5-4), HCN(1-0, $F$=2-1), HNC(1-0), H$_2$CO(3$_{1,2}$-2$_{1,1}$), and
SiO(2-1, $v=$0). The intensity values are given in units of the main beam temperature (Kelvin). On
every spectrum the corresponding IRAM beam size is indicated in arcseconds. We marked by the
shorthand ``{\bf fsw}'' those measurements that have been done with the 5~MHz frequency shift,
otherwise the beam wobbling observational mode has been used. The value of the system velocity
$V_\mathrm{lsr}=5.85$~km\,s$^{-1}$ is depicted in each panel by thick vertical line.}
\label{sd}
\end{figure}

Spectra of firmly detected lines are shown in Fig.~\ref{sd}. As can be clearly seen, they
basically exhibit three different types of line profiles. The CO(2-1), C$^{18}$O(2-1), 
HCO$^+$(1-0), and CS(2-1) lines are single-peaked, centered at the system velocity
($V_\mathrm{lsr}=5.85\pm0.1$~km\,s$^{-1}$), and narrow, $<1$~km\,s$^{-1}$ (except for CO(2-1)
which is broader than $2$~km\,s$^{-1}$). In contrast, the HCN(1-0,~$F$=2-1) and
HNC(1-0) line profiles are slightly broader and apparently have a more complicated
asymmetric structure. Furthermore, HCO$^+$(2-1), HCO$^+$(3-2), DCO$^+$(2-1), CS(5-4), and
H$_2$CO(3$_{1,2}$-2$_{1,1}$) spectra show a narrow single-peaked shape and a blueshift
up to $\Delta V \approx -2$~km\,s$^{-1}$ from the system velocity, whereas SiO(2-1,~$v=$0) has
a $\sim 1$~km\,s$^{-1}$ redshift.

The reason for such a diversity in the observed line profiles is not completely understood. We suppose
that the blueshifted single lines could actually be broad double-peaked asymmetric
spectra, with their less intense redshifted wings being lost in the noise. As we show below, all these lines but 
CO(2-1) are optically thin and their double-peaked appearance would imply that they trace mainly the rotating 
AB Aur disk but not much of the quiescent envelope material, whereas the asymmetry of the line profiles 
can be explained by the $\la5\arcsec$ pointing errors of the IRAM 30-m antenna during the measurements.
This idea is partly confirmed by the fact that some of these spectra, like HCO$^+$(3-2) and CS(5-4), require high 
densities for the excitation and were measured with small beam sizes of about $10\arcsec$ (which is comparable with the 
$\sim 6\arcsec$ disk). 

On the other hand, the IRAM beam for the HCO$^+$(1-0) and CS(2-1) lines is $\ga25\arcsec$, a few times larger than
the apparent disk size, and due to a significant beam dilution in this case, both emission lines come from the 
extended envelope around AB Aur. This suggestion is further supported by the narrow $\sim1$~km\,s$^{-1}$
width and central position of these emission lines, typical of cool and quiet gas.
This is certainly true also for the CO(2-1) and C$^{18}$O(2-1) lines, though the broad width of the former
spectrum indicates that the CO(2-1) emission is contaminated by moving gas along the line of sight to AB Aur.

We find strong evidence for a large reservoir of cold gas located within $\sim 8\arcmin$ 
around the central star that is indirectly probed by our DCO$^+$(2-1) measurements (see Table~\ref{lines},
footnote ``c)''). 
The negative intensity of this line was detected only in the case of the on-off
observations, which can be easily explained. It is known that deuterium fractionation proceeds most efficiently
in cold environments, $T\la70$~K, \citep[e.g.,][]{Bacea03} but not in the inner (warm) parts of
protoplanetary disks \citep[see][]{deut}. Consequently, the DCO$^+$(2-1) signal is stronger for the cold
outer region of the envelope at a distance of $\sim 35\,000$~AU ($240\arcsec$ offset) than for the warm inner 
$\la 2\,500$~AU part ($17\arcsec$ IRAM beam). Therefore, the measured intensity of this emission line became negative 
after the on-off subtraction of the strong background signal. Note that the HCN(1-0,~$F$=2-1) single-peaked asymmetrical 
spectrum was observed with the same $240\arcsec$ beam wobbling technique and may suffer from a similar contamination.  

It is worth to mention that an extended $\sim 35\,000$~AU region of thermal emission from cold dust grains around AB Aur 
has been observed by the {\it IRAS} satellite at $60\mu$m \citep{Whea91}. 

\subsection{The Plateau de Bure interferometric data}
\label{pdbi}
We mapped the AB Aur system in the HCO$^+$(1-0) line at $89.18$~GHz with the PdBI. These observations were obtained with 
five 15-m antennas using the compact configuration CD (baselines of 20-80 meters) in March 2002. The phase reference center
of our measurements was the same as given in the previous section.

We applied one correlator unit with a total bandwidth of only 10~MHz in order to achieve high frequency resolution of 
$\approx 0.12$~km\,s$^{-1}$. The HCO$^+$(1-0) line was centered at the system velocity $V_\mathrm{lsr}=5.85$~km\,s$^{-1}$.
The underlying continuum was detected with broad bands only and has a very low intensity.

The band pass and phase calibration were performed on the objects CRL~618, 0528+134, and 0415+3790. Maps of $128\times128$ 
square pixels with $1\arcsec$ pixel size were produced by the Fourier transformation of the calibrated visibilities using 
natural weighting. The synthesized HPBW (Half Power Beam Width) size of the beam was $6.76\arcsec \times 5.09\arcsec$ 
($=970~\mathrm{AU} \times 730$~AU at 144~pc) with a position angle of $94\degr$. For the data reduction and final phase 
calibration, we applied the Grenoble Software Environment GAG.

We checked the observed HCO$^+$(1-0) spectra by adding the corresponding single-dish IRAM data as missed zero-spacing 
information to the interferometric visibilities. Different weightings were used, but finally it turned out that the 
appearance of the interferometric spectral map did not change much. Also, we estimated that only $\sim 20\%$ of the 
flux is recovered by utilizing the zero-spacing data. Therefore, in our modeling we use the HCO$^+$(1-0) PdBI data without 
zero-spacing correction.

\begin{figure}
\includegraphics[width=0.4\textwidth,angle=270,clip=]{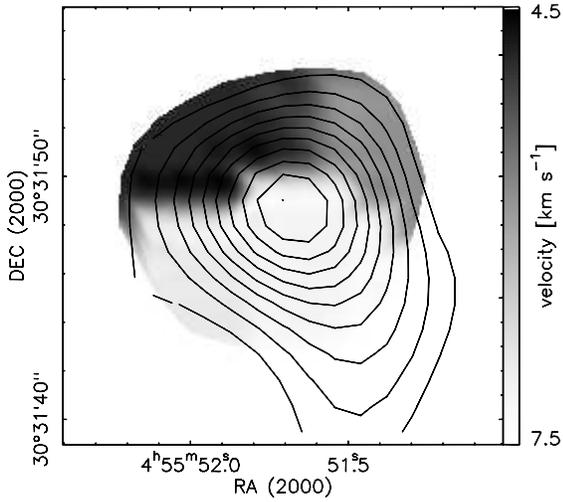}
\caption{Smoothed velocity map (in km\,s$^{-1}$) of the AB Aur system observed
with the Plateau de Bure interferometer in the HCO$^+$(1-0) transition at $\sim 89$~GHz (synthesized HPBW beam 
size is $6.76\arcsec \times 5.09\arcsec$). The distribution of the
integrated line intensity within this $\sim10\arcsec$ object is represented by contour lines (green in electronic
edition). The levels are equidistant with $0.05$~K\,km\,s$^{-1}$ intensity steps. The ``butterfly''
(``blue-red'' in electronic edition) symmetric appearance of the map is characteristic of a
globally rotating disk-like configuration that is observed close to face-on ($\iota \sim 0\degr$). 
The border between these two lobes corresponds to the projection of the disk rotational axis on the sky plane
($V=V_\mathrm{lsr}$). It implies that the disk positional angle $\phi$ is about $90 \degr$.}
\label{velo}
\end{figure}

As an example, in Fig.~\ref{velo} we show the distribution of the HCO$^+$(1-0) intensity-weighted velocities within the AB 
Aur system, over plotted by the integrated line intensity (contour lines). Though not well spatially resolved with our 
$\sim 6\arcsec$ beam, it appears as a nearly spherically symmetric $\sim10\arcsec$ ($1450$~AU) structure with a peculiar 
``two-lobe'' (blueshifted and redshifted) velocity pattern. This is a typical sign of a rotating disk-like configuration 
that is seen close to face-on \citep[compare with the upper right panel in Fig.~3 by][]{MS97}. The lack of spatial 
resolution
does not allow to determine the radius of this structure precisely, and only an upper limit can be put on this value, 
$R_\mathrm{disk} \la 800$~AU. The border between two velocity lobes (zero-velocity gradient, $V=V_\mathrm{lsr}$) 
corresponds to the projection of the disk rotational axis on the sky plane. This implies that the disk positional angle is 
about $90\degr$. 

Whether the disk size and orientation are indeed close to these first-order observational estimates is verified by the 
extensive modeling in the following sections.

\section{Model of the AB Aur system}
\label{mod}

\begin{figure}
\includegraphics[width=0.55\textwidth,clip=]{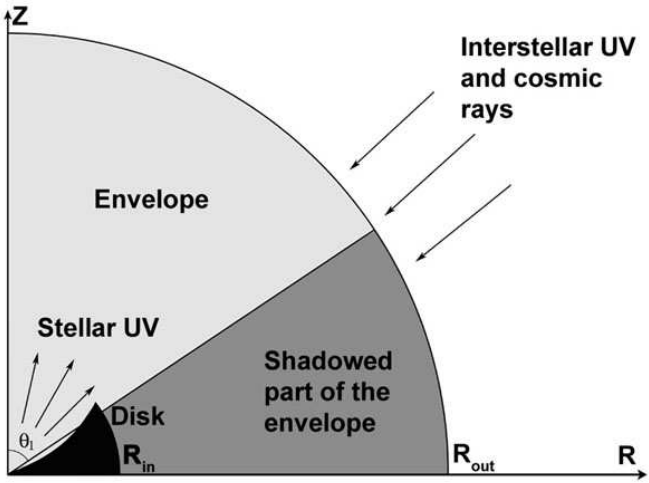}
\caption{This is a sketch of the AB Aur system (due to the symmetry of the model only one
quadrant of its vertical slice is presented). The passive flared disk is shown in black, 
while the surrounding spherical envelope is painted in gray. The envelope is further divided in the torus
region shadowed from the direct stellar light by the disk (dark gray) and two unshielded 
lobes (light gray; only one is shown). The geometrical parameters of the best-fit AB Aur model are
the following: $R_\mathrm{in}=400$~AU, $R_\mathrm{out}=2\,200$~AU, and $\theta=\pi/2-\theta_1=25\degr$.}
\label{scheme}
\end{figure}

In this section, we describe and discuss the physical and dynamical model of the AB Aur system in detail. A schematic 
sketch of this object is presented in Fig.~\ref{scheme}. Briefly, it is assumed that the AB Aur system consists of a passive flaring rotating disk surrounded by an 
extended infalling spherical envelope. This dense protoplanetary disk shades off a torus region in the diffuse envelope 
from strong stellar UV flux, which allows many complex molecules to form and survive there. In contrast, the composition of
the two unshielded envelope lobes is mainly atomic. Since the star directly heats these lobes, they have to be hotter; in
a pure hydrostatic equilibrium it would also imply that they are less dense than the shadowed part of the envelope.
Thus, an observationally significant amount of molecules is only reached during the chemical evolution of the AB Aur disk 
and shielded part of the envelope. 

\subsection{Disk structure and parameters}
\label{disk_model}
The UV-to-mm surveys of Herbig Ae stars have unveiled that many of these stars are surrounded by circumstellar 
gas and dust distributed in flattened (disk-like) configurations \citep[e.g.,][]{Nea01}. Indeed, the presence of a 
compact structure ($0.5~\mathrm{AU}\la r\la 500$~AU) around AB Aur has been revealed from IR interferometric measurements 
\citep[e.g.,][]{Mea95,MGea99,MGea01,Eea03}, at visual wavelengths by the photometric and polarimetric observations of 
\citet{GR96}, and from millimeter interferometric observations of \citet{MS97}. Moreover, the latter authors have reported 
on the Keplerian rotation of the gas in this object. All these facts indicate that a rotating protoplanetary disk encircles
the AB Aur star. Given that the measured mass accretion rate on the central star is low, 
$\dot{M}\sim 10^{-8}M_\sun$\,yr$^{-1}$ \citep[e.g.,][]{Gea96}, the vertical structure of the AB Aur disk is globally 
sustained by reprocessing stellar radiation (passive disk) but not due to viscous dissipation.

Some clues concerning the dust content of the AB Aur system can be gained from the analysis of the SED. The emission bands 
of various dust materials have been detected in the IR spectra of AB Aur with the {\it Infrared Space Observatory} (ISO), 
most notably those of iron oxide, PAHs, amorphous silicates, and water ice \citep{vdAea00}. The strong silicate emission 
band at $9.7\mu$m points to the presence of a large amount of warm ($\sim100$-$200$~K) and small ($\la0.1$-$2\mu$m) 
amorphous silicate grains in this object. Together with the absence of crystalline silicate features in the spectra, it 
implies that the dust in AB Aur is thermally unprocessed and may still resemble pristine interstellar grains even after a 
few Myr of the evolution \citep[for more details, see][]{sil, Meeusea01}.

A large amount of warm dust and gas observed in the AB Aur system \citep[e.g.][]{Tea01} cannot be accounted for 
without invoking a flared geometry of the disk or/and an additional disk heating by the surrounding envelope 
\citep[importance of the latter effect is discussed in][]{Vea03}. This idea is further supported by the results of the 
detailed modeling of \citet{Eea03} and \citet{Bea03}, who have found that a flared disk model with a puffed-up inner rim 
provides a suitable fit to their NIR- and MIR-observational data. 

Unfortunately, the disk of AB Aur is detected with the PdBI array only in one line and at modest resolution.
Thus it is not possible to unravel the disk thermal and density structure in both radial and vertical
directions, using these observational data alone. On the other hand, many other parameters of the AB Aur
system have been revealed from previous observations, which can be used to construct a disk chemical model.

In this paper, we adopt the passive flared disk model of \citet{DD03}. This disk model is calculated using a 2D 
axisymmetric continuum radiative transfer code and silicate dust opacity data \citep{DL84} under the assumption of 
hydrostatic equilibrium in vertical direction. The {\it a priori} fixed disk parameters are as follows.

We assume that the dust grains are uniform spheres with radius of 0.3$\mu$m, though the additional case of larger 1$\mu$m 
particles is also considered. In the model, we focus only on a single-size 
grain distribution instead of a range of dust sizes because otherwise it would 
significantly slow down the chemical computations and 2D disk modeling, thus
making impossible efficient fitting of the model parameters. The disk surface density as a function of radius obeys a 
power law, 
$\Sigma(r)=\Sigma_0(r/r_0)^p$, with the exponent $p=-5/2$ and the initial value $\Sigma_0$ to be constrained. Note that 
the adopted radial gradient $p$ is somewhat steeper than the usually assumed value between about $-2$ and $-1$ 
\citep[see,][]{Mea99,Nea01}. Therefore, we consider two additional models with $p=-3/2$ (minimum-mass solar
nebula) and $p=0$ (uniform disk).

The disk has its inner boundary at 0.7~AU (dust sublimation radius, $T \sim 1\,500$K) and extends up to several hundreds AU in radial 
direction, where its vertical height ranges a comparable spatial scale. The ratio of the disk outer radius to the vertical 
height at which the star is still sufficiently obscured, $A_\mathrm{V} \ga 1^\mathrm{m}$, is used to estimate what fraction
of the entire AB Aur envelope is shadowed (see Fig.~\ref{scheme}, dark gray region). 

We adopt the following parameters of the central star: $T_\mathrm{eff}=10\,000$~K, $M_{*}=2.4M_{\sun}$, $R_{*}=2.5R_{\sun}$ 
(see Table~\ref{star_par}). It is assumed that the disk is illuminated by the UV radiation from the star and by the
interstellar (IS) UV radiation. The intensity of the stellar UV flux is calculated using the \citet{kurucz} ATLAS9 of 
stellar spectra. It is converted to the standard $G$ factor, $G_{*} \approx 10^5\,G$ at the distance of 100~AU, where $G=1$
corresponds to the mean interstellar UV field of \citet{G}. To calculate the visual extinction by dust grains toward the 
central star at a given disk location, we use the following expression (1D plane-parallel case):
\begin{equation}
\label{av_to_nh}
A_\mathrm{V}={N_\mathrm{H} \over N_1(a)}~\mathrm{mag},
\end{equation}
where $N_\mathrm{H}$ is the total column density of hydrogen nuclei between the point and the star and the column density 
to reach $A_\mathrm{V}=1\,\mathrm{mag}$ is $N_1(a)=8.36\cdot10^{20}$~cm$^{-2}$ and $9.75\cdot10^{21}$~cm$^{-2}$ for the grain 
radius $a=0.3\mu$m and $1\mu$m, respectively. The extinction of the interstellar UV radiation ($G=1$) is computed in the 
same way but in vertical direction only. We calculate the penetration of cosmic rays (CR) into the disk by Eq.~(3) from 
\citet[hereafter Paper~I]{Red2} assuming an initial value of the ionization rate $\zeta_\mathrm{CR}=1.3\cdot10^{-17}$~s$^{-1}$.
Finally, ionization due to the decay of short-living radionuclides, like $^{26}$Al, is considered with a total rate 
$\zeta_\mathrm{RN}=6.1\cdot10^{-18}$~s$^{-1}$ \citep{UN81}.

\begin{figure}
\includegraphics[width=0.42\textwidth,clip=]{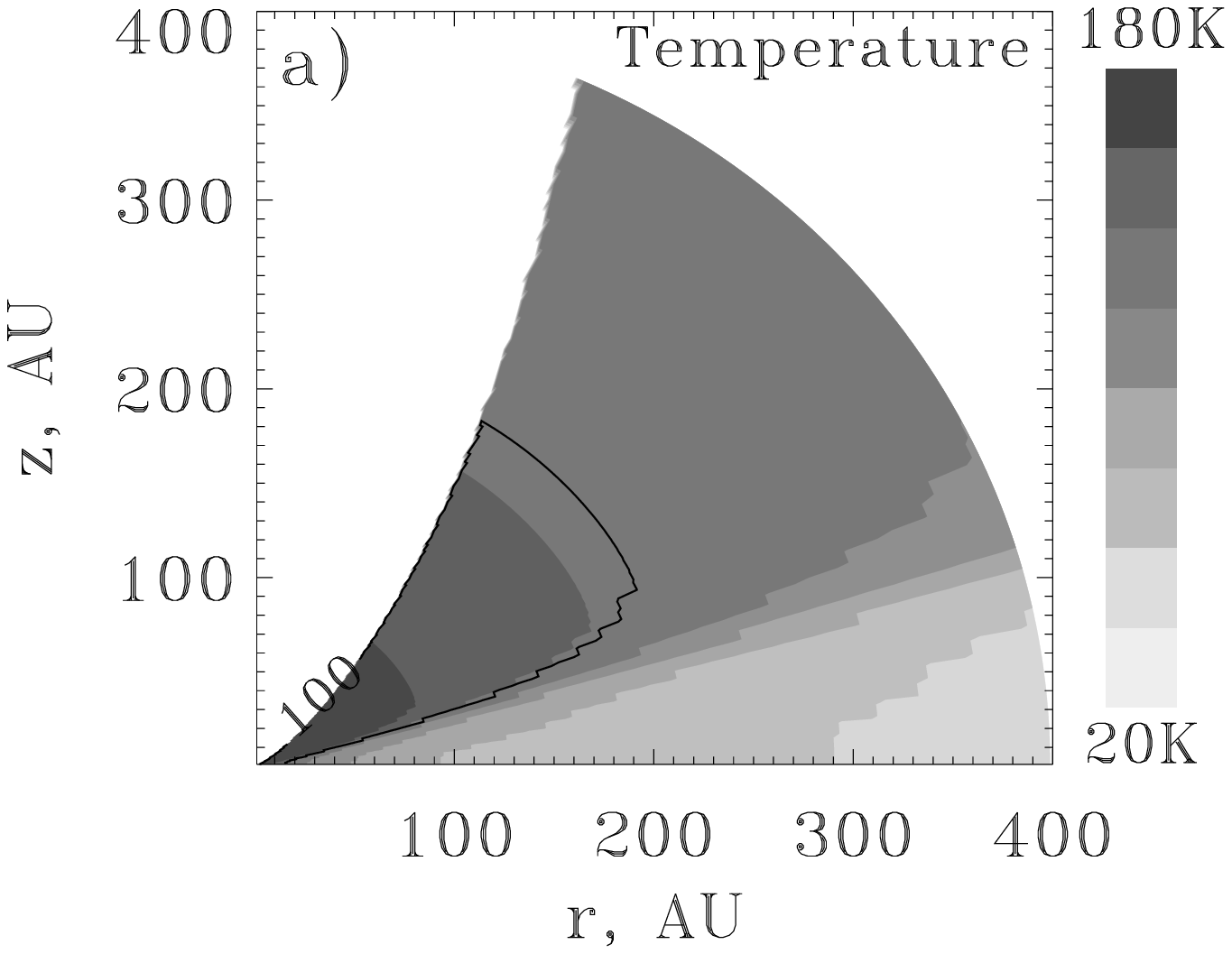}
\includegraphics[width=0.435\textwidth,clip=]{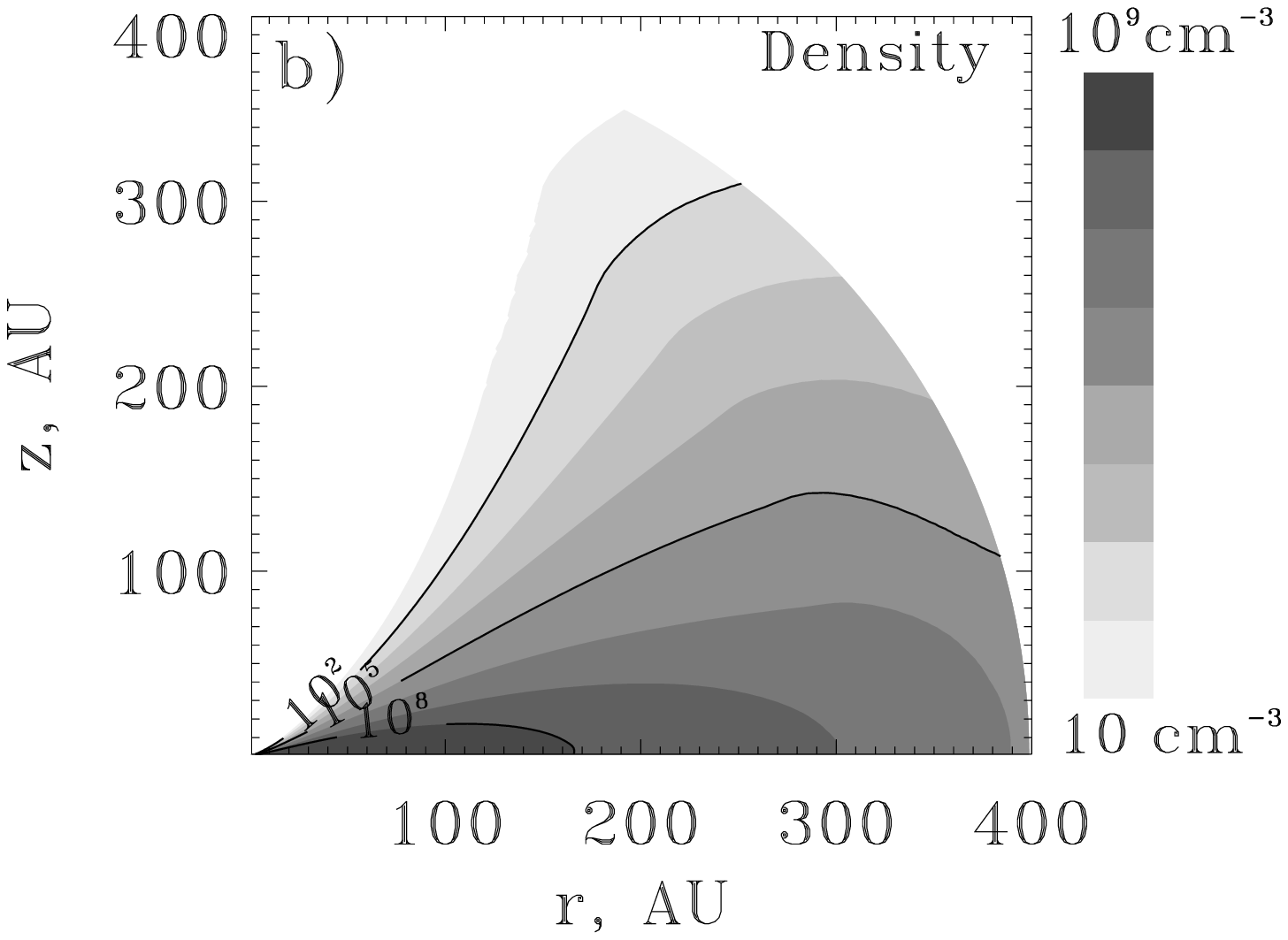}
\caption{Thermal structure of the adopted disk model is shown in the left panel, where solid
curve represents the so-called "snowline" ($T=100$~K). The relevant density distribution is presented 
in the right panel. There, three solid lines correspond to the disk regions with number densities of
$10^{2}$~cm$^{-3}$, $10^{5}$~cm$^{-3}$, and $10^{8}$~cm$^{-3}$ (from the disk surface to midplane,
respectively).}
\label{disk}
\end{figure}

Overall, a wide range of relevant physical parameters characterizes the disk model, namely, temperatures between 35~K and 
1\,500~K, densities between $10^{-21}$~g~cm$^{-3}$ and $10^{-9}$~g~cm$^{-3}$, $G$ factors $\ga10^3$, $A_\mathrm{V}$ 
between 0 and more than 100~mag, and ionization rates between $\sim 10^{-19}$~s$^{-1}$ and $10^{-17}$~s$^{-1}$. The disk thermal and density structure is shown in Fig.~\ref{disk}. 

The parameters of the best-fit disk model are given in Table~\ref{disk_par}. In Section~\ref{res} we discuss how they and 
their uncertainties are derived.

\subsection{Envelope model}
\label{env_model}
The body of observational data about the extended nebulosity surrounding AB Aur allows constraining some of its basic 
parameters prior to the detailed modeling. 

The visual extinction observed toward this star is low, $A_\mathrm{V}\sim0.2$-$0.5~\mathrm{mag}$ 
\citep{vdAea97,vdAea00,Rea01,Fea02}, and can be attributed to light scattering and absorption by dust grains in the nearby 
envelope and interstellar matter (ISM). Thus, the AB Aur envelope is diffuse and its average density has to be low.

The apparent size and morphology of the envelope depends on the spectral region at which observations are performed. For instance,
an extended $8\arcmin$ ($\approx35\,000$~AU) cloud of cold dust has been observed with the IRAS satellite at 
$60\mu$m, which is confirmed by our IRAM single-dish observations. At visual wavelengths, the spherically-symmetric envelope 
has been traced from $r\approx1\,300$~AU to $\sim 365$~AU with the HST coronographic imaging by \citet{Grea99}, who have 
also found that it is highly inhomogeneous from hundreds AU down to tens of AU (spiral arches). They have mentioned that 
this symmetric (inner) envelope is further surrounded by a large band of the reflection nebulosity that has been
detected with ground-based telescopes. 
An envelope of almost the same structure has been observed by \citet{NG95} using the {\it John Hopkins University 
Coronograph}. They have put a lower limit on the mass of the reflection nebulosity using its visual brightness, 
$M_\mathrm{refl}\sim2\cdot10^{-7}M_{\sun}$ (this seems to be a strong underestimation). \citet{DFea98} have not resolved 
the AB Aur envelope with the {\it Kuiper Airborne Observatory} at 50$\mu$m and $100\mu$m and gave an upper limit to 
its size, $r\la5\,000$~AU. 

At millimeter wavelengths, the global distribution of the circumstellar matter around AB Aur has been probed in several 
low rotational lines of CO isotopomers with the IRAM 30-m antenna by \citet{Fea02}. Fuente et al. have estimated the envelope 
mass within 0.08~pc ($\sim15\,000$~AU) to be $\sim 1\,M_{\sun}$. They have classified AB Aur as a Class~II object 
according to their notation (a star embedded in the remnant natal cloud), which implies a power-law radial density profile 
with an $-2<p<-1$ exponent. With these two limiting cases of the density profile, the derived mass 
$M_{0.08\mathrm{pc}}=1M_{\sun}$, and assuming that the envelope extends down to the dust sublimation radius $r_0=0.7$~AU, 
we find that the typical density at $r_0$ is between $10^8$ and $10^{12}$~cm$^{-3}$.

Moreover, \citet{Mea99} have fitted the SED of AB Aur using a model of a flat disk immersed in a spherically-symmetric 
diffuse envelope. Their best-fit parameters for the AB Aur envelope are the following: the mass is about $0.03M_{\sun}$,
the radial density distribution follows a broken power law, $p=-2$ for $1.2~\mathrm{AU}<r<120$ and $p=0$ for 
$120~\mathrm{AU}<r<5800$~AU with initial density of $\sim 10^8$~cm$^{-3}$, and the temperature scales with radius as 
$T(r)=1\,500~\mathrm{K}\cdot(r/1.2~\mathrm{AU})^{-0.4}$. 

Recently, \citet{Eea04} have fitted the AB Aur SED from UV to the radio domain with a similar model of a pure spherical envelope
and assuming large porous grains (dust emissivity $\propto \lambda^{-0.6}$). They have adopted radial power-law 
temperature ($q=-0.4$) and density ($p=-1.4$) distributions with $T_0=1\,500$~K and $n_0=3\cdot10^9$~cm$^{-3}$ at 
$r_0\approx1$~AU.

Using these observational facts and theoretical constraints, we construct a model of the AB Aur envelope. 
Since we probed the molecular content of the AB Aur envelope in several lines with the IRAM 30-m antenna,
it allows to derive the temperature and density structure of the envelope {\em directly} from 
these data. Thus, extensive modeling of its physical structure with an advanced radiative (hydro-)
code can be omitted.

We focus on the inner part of the entire AB Aur envelope and assume that it is spherically symmetric, homogeneous, and has 
a radius of $R_\mathrm{out}=2\,200$~AU (restricted by the largest $29\arcsec$ IRAM beam size used in our single-dish 
observations). Given a low inclination of the disk inferred from the interferometric data, the shielded part of the 
envelope is also seen close to face-on. In this case, it makes no sense to consider the outer regions of the envelope
beyond 2\,200~AU since 
they are molecularly deficient along the line of sight (apart from H$_2$ and CO), while in the opposite direction the 
emitting material is out of the used IRAM beams. We assume that the inner radius of the AB Aur envelope coincides with the 
disk outer edge, $R^\mathrm{env}_\mathrm{in}=R^\mathrm{disk}_\mathrm{out}\ga200$--600~AU. 

The envelope radial density profile is modeled with a power law, $\rho(r)=\rho_0\cdot(r/r_0)^{-p}$, where $\rho_0$ and $p$ are
parameters to be constrained. We assume that the shadowed part of the AB Aur envelope is isothermal, than its mean kinetic 
temperature $T_\mathrm{kin}$ can be directly estimated from the $T_\mathrm{mb}$ intensity of the optically thick CO(2-1) 
line: $T_\mathrm{kin} \sim T_\mathrm{mb}\approx 20$-$40$~K (see Table~\ref{lines}). Our initial guess for the inner envelope 
temperature is similar to the range $T=20$-$50$~K derived by \citet{Tea01} from the analysis of excitation conditions for 
several CO low- and high-rotational lines. We assume that gas and dust species are in thermal equilibrium, which is a 
justified assumption for the inner (dense, cold, and shadowed) part of the envelope. 

The gas-grain and grain-grain collisional timescales in the AB Aur envelope are larger than those in the dense disk.
This essentially rules out the possibility that a profound grain growth up to cm-sized bodies occurred there during the 
contraction phase, which is further supported by the fact that the bulk of dust grains in the AB Aur system are smaller than 
a few microns. Therefore, we assume that dust particles in the envelope are small uniform spheres with radius of $0.1\mu$m.

The approach of the previous subsection is adopted to compute the UV flux in the shadowed part of the envelope, but with 
some minor modifications. As this region is shielded from the direct stellar radiation, it is only necessary to take into 
account the penetration of the IS UV photons. Given the low radial density of the envelope ($\Sigma_\mathrm{env} \ll 
100$~g~cm$^{-2}$), the constant CR-ionization rate is taken, $\zeta_\mathrm{CR}=1.3\cdot10^{-17}$~s$^{-1}$, whereas ionization
due to the decay of the radionuclides is negligibly low, $\zeta_\mathrm{RN}=0$.

The best-fit parameters of the AB Aur envelope model are compiled in Table~\ref{env_par} and discussed in Section~\ref{res}.

\subsection{Chemical model}
\label{chemical_model}
The chemical model adopted in this study is the same as in Paper~I, but with a few modifications. Briefly, we use the 
UMIST\,95 database of gas-phase reactions \citep{umist95} supplied by a set of dust surface reactions from 
\citet{HHL92} and \citet{HH93}. In the rates of reactions with cosmic ray particles and CR-induced UV photons the 
CR-ionization rate, $\zeta_\mathrm{CR}$, is replaced either by the sum of $\zeta_\mathrm{CR}+\zeta_\mathrm{RN}$ (disk) or 
left unchanged (envelope). 
Contrary to Paper~I and in accordance with the 
studies by \citet{WL00}, \citet{Aea02}, and \citet{vZea03}, we assume that the probability of species to stick onto dust 
grain surfaces is 100\%. Finally, we enlarge the chemical network with a set of deuteration reactions (Bergin, personal 
communication) and take into account self- and mutual-shielding of H$_2$ and CO molecules. Our chemical network does 
not include reactions involving C$^{18}$O, therefore we scale down the CO abundances calculated without self-shielding
by the isotopic ratio O/$^{18}$O=490 \citep{WR94}.

Overall, this network consists of 560 species made of 13 elements, and 5336 reactions.

Though \citet{IHMM04} have found that the vertical mixing and radial transport may significantly affect the disk chemical 
evolution under certain conditions, we do not consider these processes in our model because of three main reasons. First, the 
estimated mass accretion rate on AB Aur is low, therefore the radial transport of matter toward the central star 
is slow. Second, there is observational evidence that turbulence in protoplanetary 
disks is low, $V_\mathrm{turb} \la 0.1$-0.2~km\,s$^{-1}$ \citep[see, e.g.][]{DDG03}, and consequently 
diffusion processes can be not that important. Last (but not least) reason why mixing processes are not considered is that 
the chemical model has to be remain the numerically manageable.

\subsubsection{Deuterium chemistry}
We adopt deuterium chemical reactions from \citet{BNM99}. This set includes essential formation and destruction routes relevant 
to the chemical evolution of OD and HDO. It consists of about 60 gas-phase and 10 gas-grain reactions among 11 
species including accretion onto and desorption from dust surfaces of D, OD, and HDO species. The elemental abundance of 
deuterium, D/H=$1.52\cdot10^{-5}$, is taken from \citet{Pea97}.

\subsubsection{Self- and mutual-shielding of H$_2$ and CO}
The photodissociation of H$_2$ occurs through discrete absorption in Werner and Lyman bands in the wavelength range 
between $912$~{\AA} and $1\,110$~{\AA} with an efficiency of $\sim 10\%$ \citep[e.g.,][]{DS70}. Given the fact that these
lines become optically thick when the hydrogen column density reaches $\sim 10^{14}$~cm$^{-2}$, it is unavoidable to take 
into account the effect of the H$_2$ line self-shielding both for the disk and envelope chemistry. To calculate the H$_2$ 
shielding, we use an analytical expression~(37) from \citet{DB96},
\begin{equation}
\label{h2ss}
F_\mathrm{sh}(\mathrm{H}_2)=
\frac{0.965}{(1+x/b_5)^2}+\frac{0.035}{(1+x)^{0.5}}\,\exp\left[-8.5\cdot10^{-4}\,(1+x)^{0.5}\right],
\end{equation}
where $x=N(\mathrm{H_2})/5\cdot10^{14}$~cm$^{-2}$ and
$b_5=3\,\sqrt{T/100~\mathrm{K}}$~cm~s$^{-1}$. The unattenuated dissociation rate of hydrogen molecules for the interstellar
UV radiation ($G=1$) is taken as $3.4\cdot10^{-11}$~cm$^3$~s$^{-1}$. 

The photodissociation of CO molecules is dominated by discrete absorption at $\lambda\la1120~\mathrm{\AA}$, 
though it also proceeds via continuum absorption \citep{vD88}. Since hydrogen molecules dissociate in the same wavelength 
range, one has to account for mutual-shielding by coincidental lines of H$_2$ and dust extinction when calculating the CO 
self-shielding. This is a complex task and there is no a convenient analytical representation similar to Eq.~(\ref{h2ss}). 
Therefore, we follow the approach of \citet{vZea03} and compute the CO self-shielding by interpolating the values from 
Table~11 of \citet{Lea96}, where a set of the shielding factors is given as a function of the H$_2$ and CO column densities
and visual extinction by dust grains. As initial guess for the CO column densities in the medium under consideration, we 
use those of H$_2$ and scale them by a constant factor, CO/H$_2=6\cdot10^{-5}$. The adopted unattenuated photodissociation 
rate of CO molecules is $2.0\cdot10^{-10}$~cm$^3$~s$^{-1}$.

\subsubsection{Results of chemical calculations}
\label{chem_res}
The initial abundances for the chemical model are calculated as in Paper~I. Concisely, we simulate the 
chemical evolution of a molecular cloud with uniform temperature $T=10$~K and density $n_\mathrm{H}=2\cdot10^4$~cm$^{-3}$ for 
1~Myr with the reaction network described above and the ``low metal'' abundances from \citet{WSH03}. 

Using these pre-computed quantities and the model of the AB Aur system, a set of time-dependent molecular abundances and 
vertical column densities are computed for 3~Myr of the evolution in every iterative step of the modeling. Below we discuss 
the results obtained with the final (best-fit) set of model parameters (Tables~\ref{star_par}--\ref{env_par}).

\begin{figure}
\includegraphics[width=0.44\textwidth,clip=]{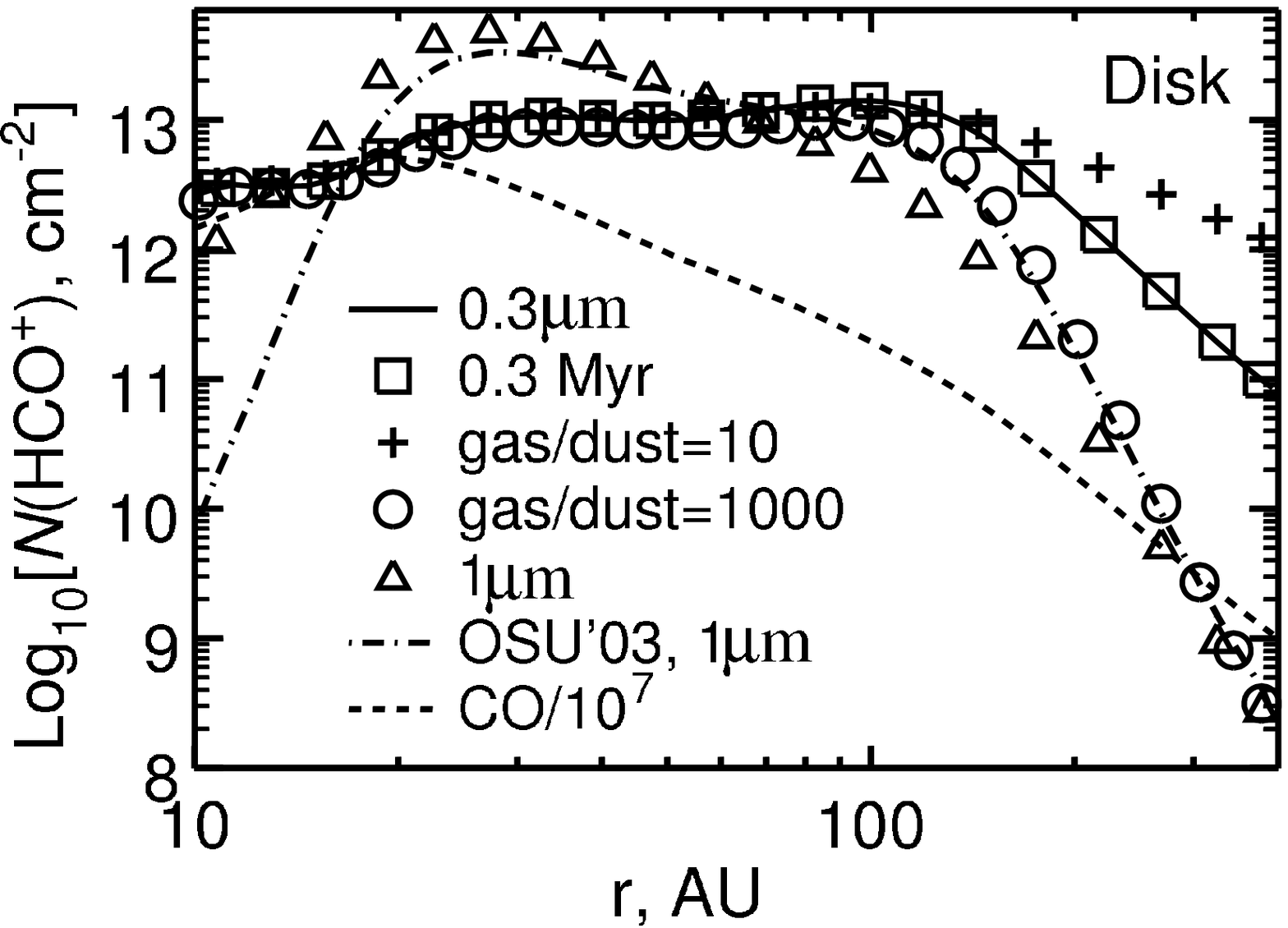}
\includegraphics[width=0.43\textwidth,clip=]{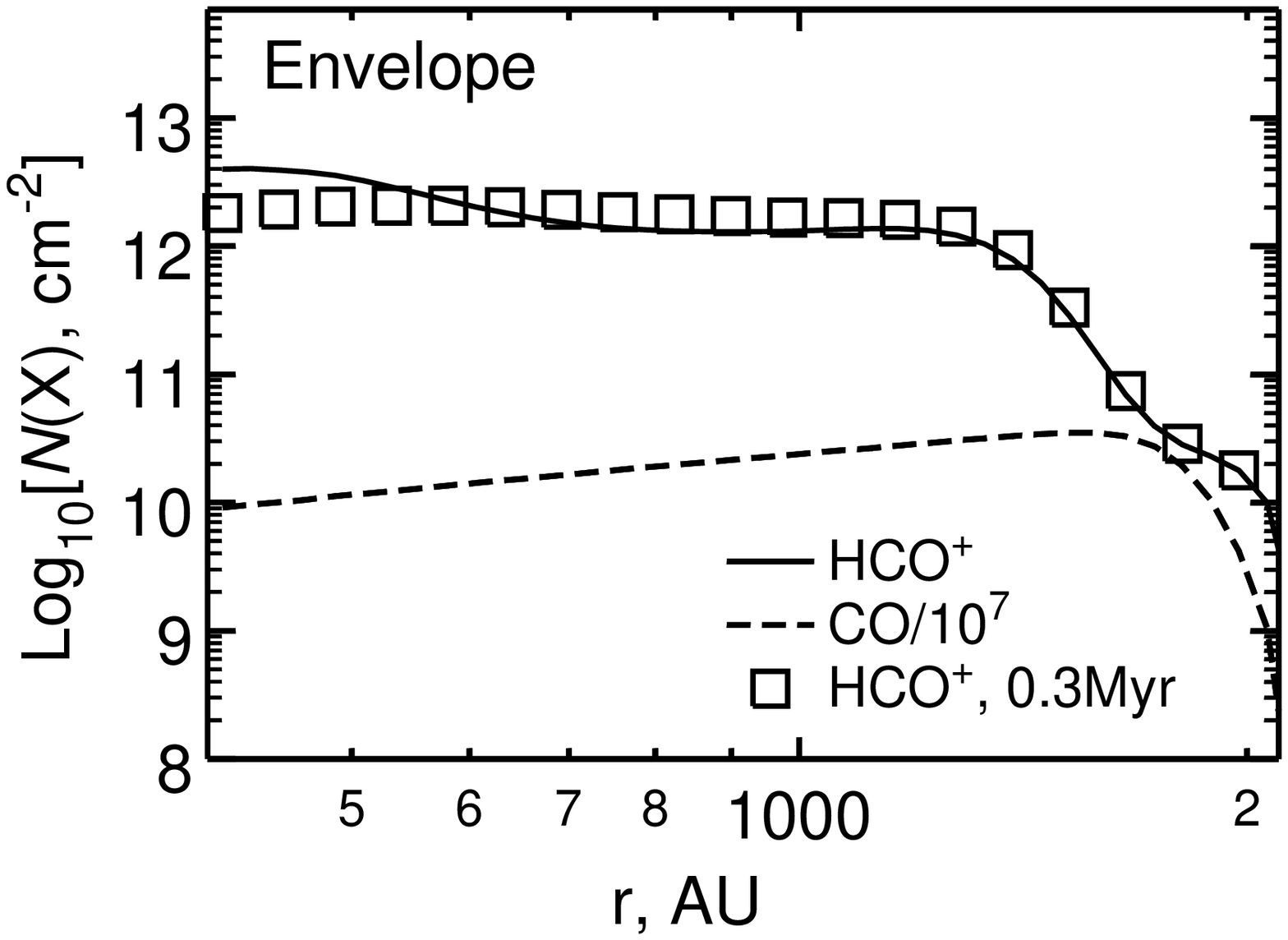}
\caption{Radial distributions of the calculated vertical column densities in the disk (left panel) and
envelope (right panel). Thick solid line represents the results for HCO$^+$ obtained with the standard model 
of the UMIST\,95 gas-grain chemistry assuming $0.3\mu$m grains and 3~Myr evolutionary time span, whereas open squares 
correspond to the case of earlier $0.3$~Myr chemistry. The same standard model is used to compute column densities for 
CO molecules, which are depicted by dashed lines. Note that the CO column densities are scaled by a factor of $10^7$. In the 
left panel, we denote the HCO$^+$ disk column densities calculated with the model of larger $1\mu$m dust grains by open
triangles, whereas crosses and open circles stand for the results obtained with the standard model, but two different 
gas-to-dust ratios, 10 and 1000, respectively. Finally, dashed-dotted line in the left panel corresponds to the 
HCO$^+$ column densities in the disk after 3~Myr of the evolution, which are computed using the OSU\,03 gas-grain chemical 
network with $1.0\mu$m grains.}
\label{colden}
\end{figure}

The calculated HCO$^+$ and CO vertical column densities as a function of radius are shown in Fig.~\ref{colden}. 
As can be clearly seen, $N_\mathrm{CO}(r)$ in general follows the radial decline of the surface density and thus is less 
steep for the AB Aur envelope than for the disk. The fact
that CO column densities are strongly related to the total amount of hydrogen in both the disk and envelope, 
$N$(CO)/$N$(H)$\la10^{-4}$, merely reflects chemical stability and efficient shielding of CO molecules from 
dissociating UV radiation. In contrast, the HCO$^+$ column densities as a function of radius show a more complicated behavior
because the chemical evolution of this molecular ion is governed by a larger set of formation and destruction pathways 
(compare solid and dashed lines in Fig~\ref{colden}). For instance, a rapid decrease of $N_{\mathrm{HCO}^+}$ in the outer 
parts of the disk and envelope is caused not only by the decline of the surface density, but also due to enhanced 
recombination rate of HCO$^+$ molecular ions with abundant electrons. These regions are rather transparent for the impinging
UV photons that dissociate molecules and ionize chemical species (mostly carbon atoms), thereby increasing the electron 
concentration in the gas-phase and decreasing the HCO$^+$ abundances.

The difference between the chemical evolution of the disk and shielded part of the envelope is the following. The disk has 
density, temperature, and UV-intensity gradients in both vertical and radial directions, leading to a ``layered'' chemical 
structure. The abundances of many gas-phase species peak at the disk intermediate layer where the UV flux is mild 
enough to drive a rich molecular chemistry \citep[e.g.,][]{Aea02,Red2}. Contrary to the disk, the shadowed envelope 
region is cold, less dense, and opaque to the stellar UV radiation (but not to the interstellar UV photons). Therefore, 
many photoreactions and reactions with barrier cannot efficiently proceed there, which results in a more simplified 
(``dark'') chemistry.
  
It is interesting that the chemical equilibrium for HCO$^+$ and many other species is achieved almost everywhere in the 
disk and envelope at $t\ga0.1$-$1.0$~Myr (compare open squares and solid line in the left and right panels in 
Fig.~\ref{colden}). However, we find that this is not true for deuterated species whose abundances and column densities 
vary during the entire 3~Myr of the evolution, and thus may serve as a ``chemical clock'' \citep[e.g.,][]{deut}.
We also investigate how various physical and chemical factors influence the evolution of HCO$^+$ in the disk in detail. 

First, we 
consider the standard model of the UMIST\,95 chemistry, $0.3\mu$m grains, and $t=3$~Myr, but two limiting gas-to-dust mass 
ratios, 10 and 1000. As can be clearly seen, the resulting column densities are nearly the same as in the case of
the standard value $m_\mathrm{gd}=100$ for $r\la100$~AU, $N_{\mathrm{HCO}^+}(r) \sim 10^{13}$~cm$^{-2}$ (compare crosses, 
open circles, and solid line in Fig~\ref{colden}, left panel). At larger radii, these quantities for $m_\mathrm{gd}=1000$ 
decrease faster with radius than the column densities computed with the standard chemical model and the difference at 
$r=400$~AU reaches about 3 orders of magnitude. Contrary, in the case of $m_\mathrm{gd}=10$ the radial decline is shallower 
than in the standard case, and at $r=400$~AU the calculated column densities differ by a factor of 10. A similar 
tendency is seen for the model of larger 1$\mu$m dust particles. In the inner disk regions, $13~\mathrm{AU}<r<60$~AU, the 
corresponding HCO$^+$ column densities are $\la 10$ times higher than that of the standard model $m_\mathrm{gd}=100$, while
at larger radii they decrease with radius like in the case of $m_\mathrm{gd}=1000$ (compare solid line with open triangles 
in the left panel in Fig~\ref{colden}). 

All these trends can be understood if one recalls the fact that the disk at $r\ga100$~AU has a so low surface density
that the interstellar UV radiation can easily penetrate and thus controls the chemistry. As we noted above, the 
HCO$^+$ abundances strongly depend on the total amount of electrons in the gas-phase. In turn, the overall electron 
concentration in the outer disk regions is related to the value of dust extinction in vertical direction: the higher the 
extinction is, the lower is the electron concentration (and the higher is the resulting HCO$^+$ column densities). For a fixed 
hydrogen column density and two models of uniform grains, the ratio of the corresponding visual extinction values is 
\begin{equation}
\frac{A_\mathrm{V,1}}{A_\mathrm{V,2}}=
\frac{m_\mathrm{gd,2}}{m_\mathrm{gd,1}}\,\frac{a_\mathrm{d,2}}{a_\mathrm{d,1}}\,
\frac{Q\mathrm{ext}(a_\mathrm{d,1},\mathrm{V})}{Q\mathrm{ext}(a_\mathrm{d,2},\mathrm{V})},
\label{Av_ratio}
\end{equation}
where $m_\mathrm{gd}$ is the gas-to-dust mass ratio and $Q_{\mathrm{ext}}(a_\mathrm{d},\mathrm{V})$ is the extinction
efficiency factor for a particle with the radius $a_\mathrm{d}$.

According to this equation, the disk with the gas-to-dust ratio of $1000$ is 10 times more transparent in vertical
direction than in the case of the standard model with $m_\mathrm{gd}=100$, while the disk with 10 times more dust grains 
($m_\mathrm{gd}=10$) is 10 times more opaque. Similarly, in the case of larger $1\mu$m grains, the total visual 
extinction is $\approx6$ times smaller than that of the standard $0.3\mu$m grain model due to $\approx6$ times higher 
extinction efficiency and about 27 times reduced amount of dust particles in unit volume. Moreover, the $\sim 3$ times 
larger radius of $1\mu$m dust grains in respect to the $0.3\mu$m particles also implies that Coulomb attraction of 
positively charged ions and negatively charged dust particles is smaller, leading to less effective grain neutralization 
of HCO$^+$ ions. In turn, this results in the bump of the HCO$^+$ column density curve at $13~\mathrm{AU}<r<60$~AU 
(compare solid line with open triangles in Fig~\ref{colden}).

Another important question that needs to be clarified is how sensitive the chemical modeling to the adopted set of 
reactions is. There are two available chemical databases that are widely used, namely,
UMIST\footnote{\url{http://vizier.cfa.harvard.edu/viz-bin/VizieR?-source=J/A+AS/121/139}}\,95 
\citep[University of Manchester, T. Millar's group;][]{umist95} and 
OSU\footnote{\url{http://www.physics.ohio-state.edu/$\sim$eric/research\_files/cddata.july03}}\,03 
\citep[Ohio State University, E. Herbst's group;][]{OSU03}. These networks comprise hundreds of species 
and thousands of reactions with the rates that are not always the same \citep{Jenam2003}.

To answer this question, we repeat the chemical calculations for the disk model with $1\mu$m dust grains using the OSU\,03 
network instead of the UMIST\,95 reaction set. The corresponding HCO$^+$ column densities are depicted in Fig.~\ref{colden},
left panel, dot-dashed curve. As can be clearly seen, $N_{\mathrm{HCO}^+}$ for the OSU\,03 and UMIST\,95 models are 
nearly the same everywhere in the disk except for the very inner parts, $r\la15$~AU (compare dot-dashed line and open 
triangles in the Figure). There, the column densities of HCO$^+$ computed with the OSU\,03 model decrease more rapidly with 
radius than in the case of the UMIST\,95 network. In a forthcoming paper (Semenov et al., in preparation), we will compare 
the results of the disk chemical modeling with the UMIST\,95 and OSU\,03 networks and investigate in detail what the 
reasons for the difference in resulting molecular abundances are.

Moreover, recently \citet{Vasyunin04} have shown that computed abundances of HCO$^+$ ions in dark and diffuse molecular 
clouds can be uncertain by a factor of $\sim1.8$-$3$ due to uncertainties in the reaction rate coefficients of the UMIST\,95 
database (see Fig.~1 therein). Given a broad range of physical conditions encountered in the applied disk model, most 
probably the calculated HCO$^+$ column densities suffer from the same uncertainty of a factor of $\sim2$.

The total computational time for the disk model with 7 radial and 11 vertical grid points is $\sim 10$ hours on a Pentium IV 
2.4~GHz PC, while it takes about several hours to compute the chemical evolution of the envelope model with 28 grid points.

\section{2D line radiative transfer calculations}
\label{lrt}
In this section, we describe our approach to simulate the radiative transfer in molecular lines and to synthesize their 
spectra. In essence, the radiative transfer modeling is based on the solution to the radiative transfer (RT) equation 
coupled with balance equations for molecular level populations \citep[ see, e.g.,][]{RH91}. Prior to the modeling, one needs 
to provide density, temperature, and velocity distributions as well as molecular abundances in the medium. 

We solve the system of the radiative transfer and balance equations with the 2D non-LTE code ``URAN(IA)'' developed by 
\citet{Urania}. It partly uses the scheme originally proposed and implemented in the publically available 1D 
code ``RATRAN'' \citep{HvT00}. The iterative algorithm of ``URAN(IA)'' is the following.

First, initial molecular level populations and a set of photon random paths through the model grid have to be defined. 
Using these quantities, the specific intensities $I_\nu(i,j)$ are computed for each cell $i$ by the explicit integration 
of the RT-equation along the pre-defined photon paths $\vec{n}(j)$. For the photon ray-tracing, the code employs a 
Monte Carlo description. After that, $\vec{n}(j)$ and $I_\nu(i,j)$ are used 
to calculate the mean line intensity $J_\nu(i)$ in every cell. The computed mean intensities are utilized in the next iteration
step to refine the level populations by solving balance equations in all model cells. To accelerate convergence of the
entire procedure for optically thick lines, additional internal subiterations for each grid cell (ALI scheme) are included 
on top of the global iterations. The adopted acceleration scheme relies on the fact that the calculated mean line intensity 
in every particular cell can be divided into internal component generated in the cell and external contributions from other 
cells of the grid. Therefore, subiterations are applied to bring into an agreement the internal mean intensity of the
line and corresponding level populations \citep[for more detail, see][]{HvT00}.

The global iterations are performed until the final molecular level populations are obtained. After that, we repeat the 
calculations again, but with another set of pre-defined random photon paths in order to estimate a typical error of the 
computed values. In our simulations, the relative errors in the level populations are always smaller than $5\%$. Finally, 
the resulting level populations are used to calculate the corresponding excitation temperatures, which are further 
transformed into synthetic beam-convolved single-dish and interferometric spectra. 

There are a few limitations in our code. Since ``URAN(IA)'' does not contain a realization of the radiative transfer 
in the lines with fine structures yet, formally we are only allowed to consider rotational transitions of CO, CS, HCO$^+$ 
and their isotopomers among the full set of the detected species. The utilized collisional rate data for CO, CS, 
and HCO$^+$ are taken from \citet{cr_co}, \citet{cr_cs}, \citet{cr_hcopa}, and \citet{cr_hcopb}, respectively. Also, we 
do not take into account the continuum absorption and emission by dust grains, which is a reliable approximation for 
optically thin and moderately optically thick lines \citep[e.g.,][]{LL76}.

\subsection{Calculated excitation temperatures}
\label{texc}
For the line radiative transfer modeling of the AB Aur system we use the set of input parameters compiled in 
Table~\ref{lrt_par}. Standard Keplerian rotation is adopted to represent the regular velocity field of the disk. 
The quality of the acquired interferometric data does not permit us to verify from the radial velocity shifts of the 
observed lines whether the disk rotation is indeed Keplerian ($V_\mathrm{disk}(r) \propto r^{-0.5}$) or follows another 
power law \citep[see, e.g.,][]{SDG00}. However, the AB Aur disk has a much smaller mass in comparison with the stellar 
mass \citep[e.g.,][]{MS97}, therefore such an assumption should be valid. 

Little is known about the dynamical state of the AB Aur envelope. Therefore, we consider a steady accretion of the envelope
material on the disk (and consequently on the central star), with the regular velocity that can be found from the 
conservation-of-mass principle and adopted power-law exponent $p$ of the radial density distribution: 
$V_\mathrm{env}(r) \propto r^{-2-p}$. For example, for $p=-1$ the infall velocity decreases as 
$V_\mathrm{env}(r) \propto r^{-1}$. In agreement with recent observations by \citet{Tea01} and \citet{DDG03}, the uniform 
microturbulent velocity $V_\mathrm{turb}=0.2$~km\,s$^{-1}$ is assumed for both the disk and envelope model.

\begin{figure}
\begin{center}
\includegraphics[width=0.44\textwidth,clip=]{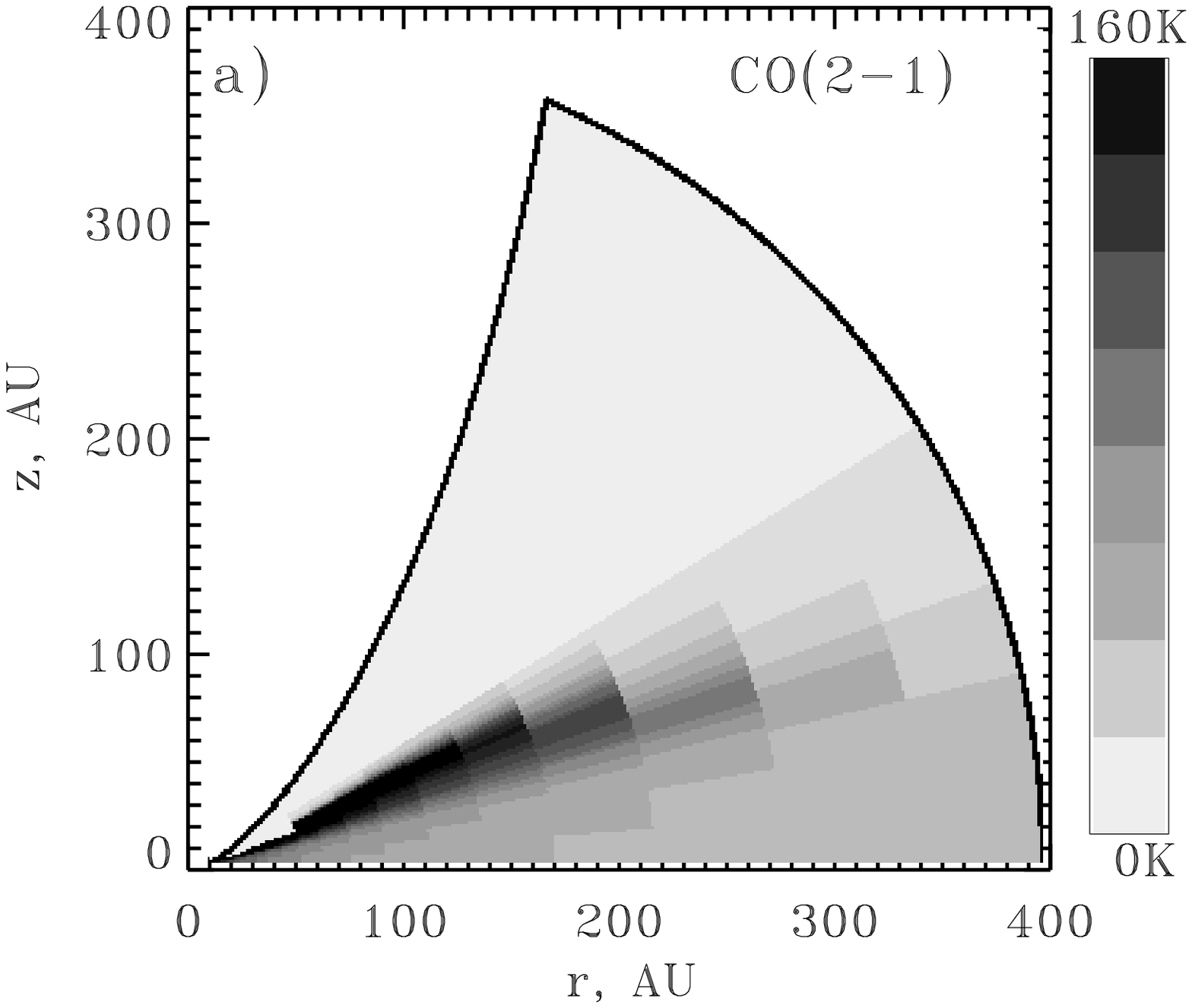}
\includegraphics[width=0.44\textwidth,clip=]{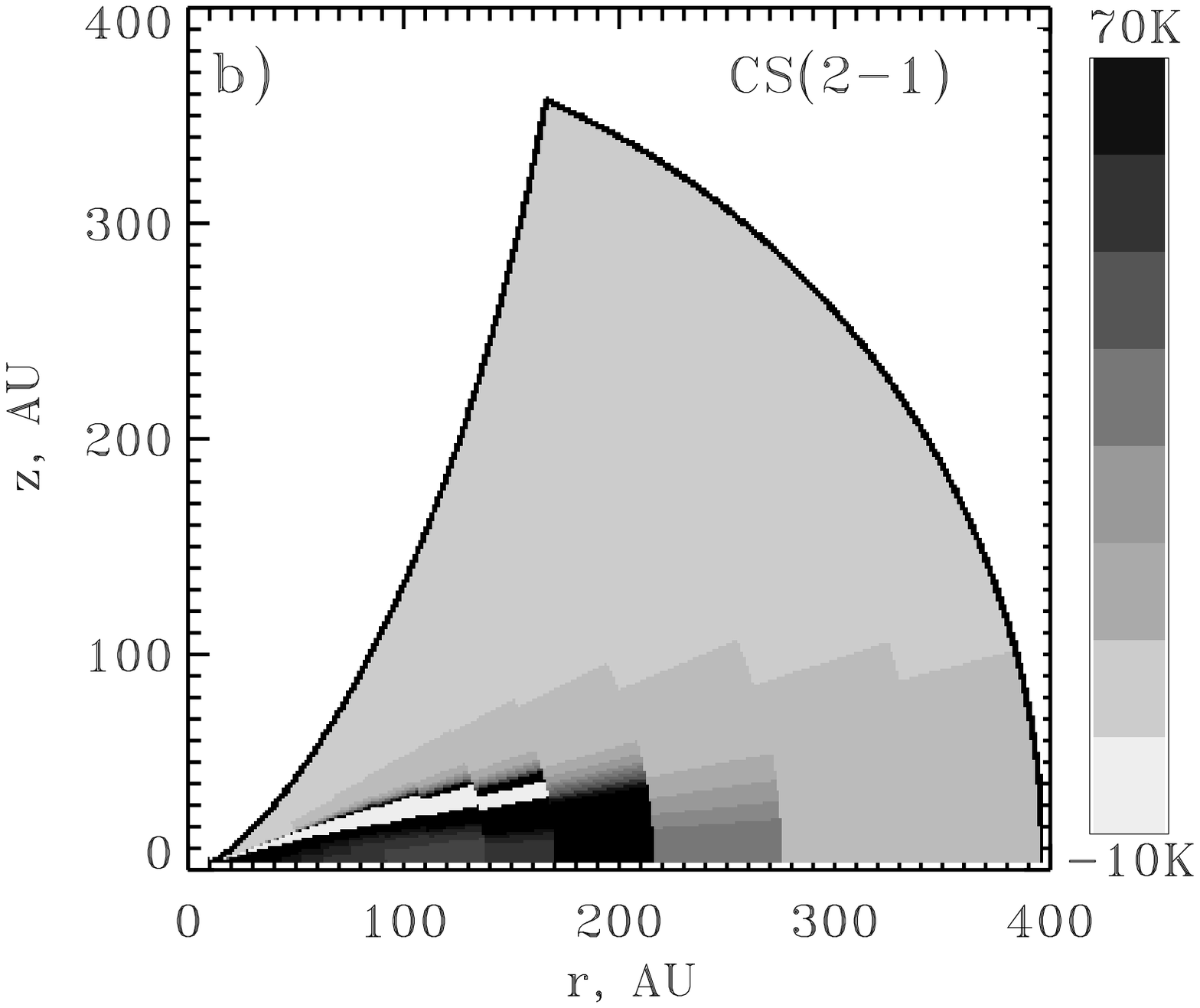}
\includegraphics[width=0.44\textwidth,clip=]{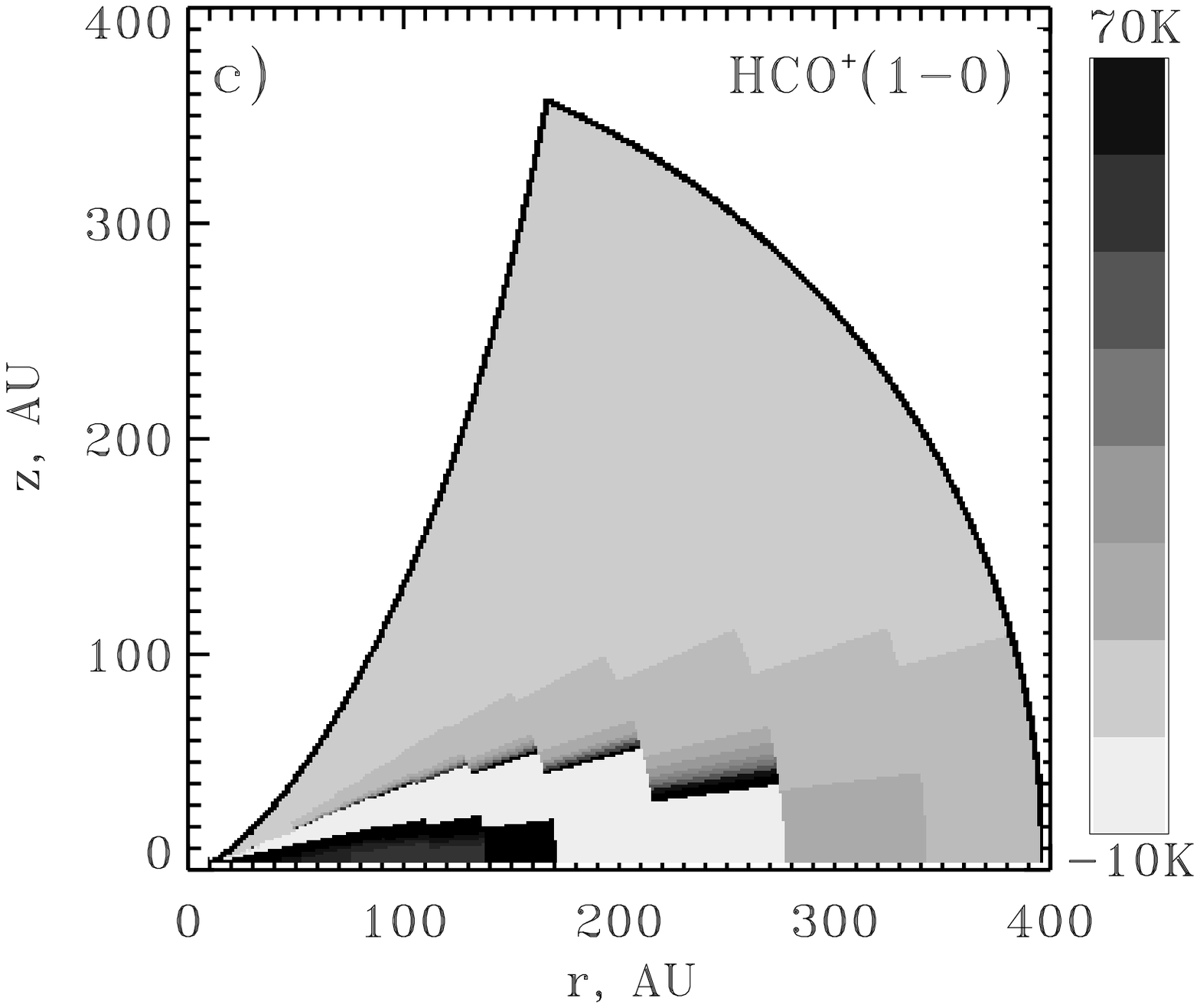}
\includegraphics[width=0.44\textwidth,clip=]{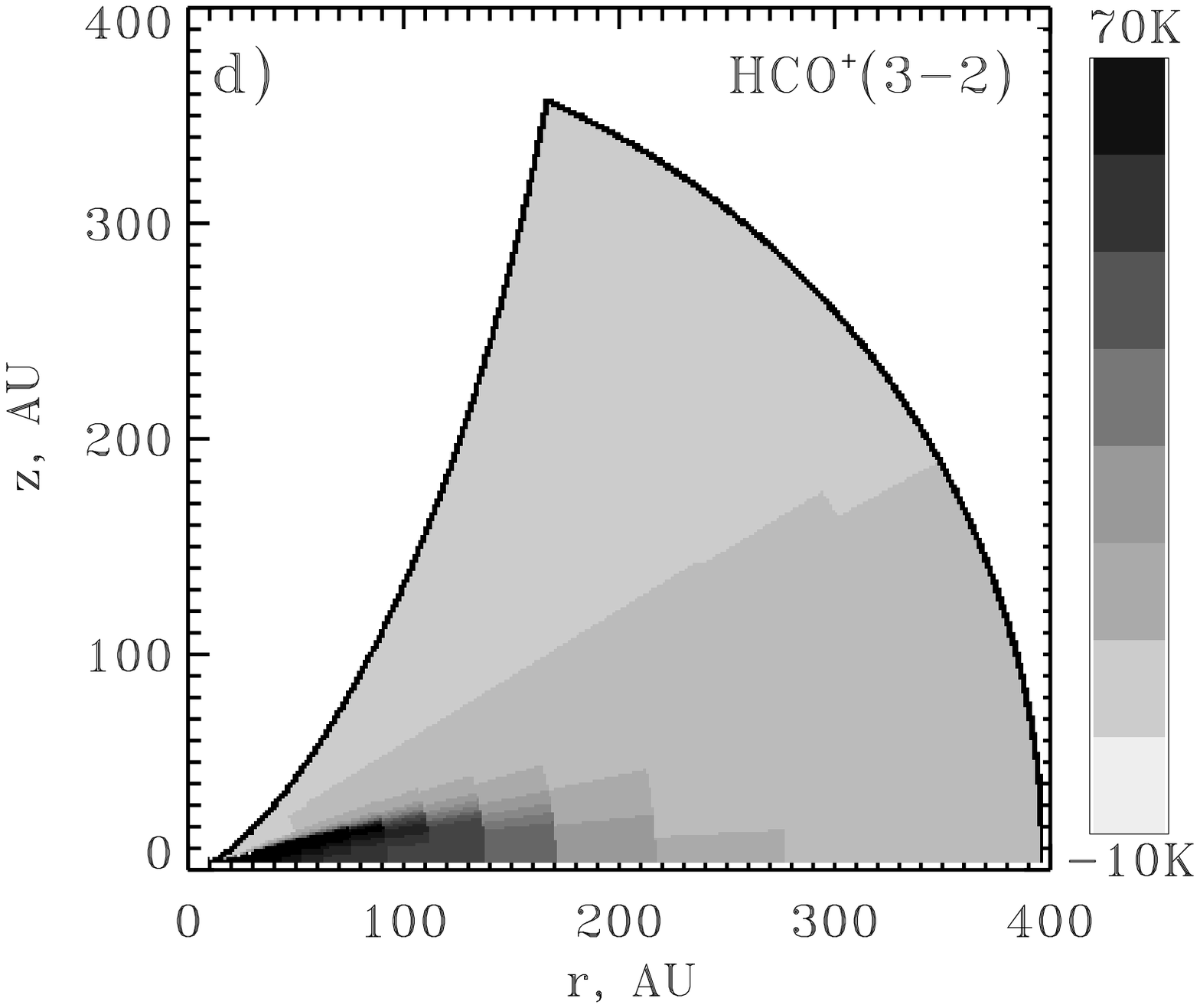}
\end{center}
\caption{Calculated excitation temperatures in the disk for the CO(2-1), CS(2-1), HCO$^+$(1-0), 
and HCO$^+$(3-2) transitions are presented in the upper left ({\bf a}), upper right ({\bf b}), 
lower left ({\bf c}), and lower right ({\bf d}) panels. Light gray areas in the panels ({\bf b}) and
({\bf c}) indicate the disk regions with negative excitation temperature (inversion in the level
populations). The plots look somewhat coarse due to a finite discretization of the applied 
disk model.}
\label{ex}
\end{figure}

The computed disk excitation temperatures of the CO(2-1), CS(2-1), HCO$^+$(1-0), and HCO$^+$(3-2) transitions are shown in 
Fig.~\ref{ex}. As can be clearly seen, the excitation temperature $T_\mathrm{ex}$ for the CO(2-1) line follows the kinetic 
temperature everywhere in
the disk but the surface layer (compare the left panel in Fig.~\ref{disk} with the top left panel in Fig.~\ref{ex}). This line is 
easily excited (critical density $n_\mathrm{cr}\sim10^3$~cm$^{-3}$) and thus thermalized in most of the disk and envelope 
regions, $n_\mathrm{H}\ga10^5$~cm$^{-3}$. Given this fact and the high abundance of CO molecules, CO/H$_2\la10^{-4}$, the 
CO(2-1) line is optically thick with $\tau \sim 10$-$1000$ in the disk and $\tau \la 10$ in the envelope.

In contrast, the CS(2-1) transition requires about 200 times higher density for thermalization, therefore its 
excitation temperature peaks closer to the dense disk interior (see the top right panel in Fig.~\ref{ex}). The optical 
thickness of the CS(2-1) line is low, $\tau\sim10^{-3}$, mainly due to much lower gas-phase CS abundances compared with 
the amount of CO molecules, $N(\mathrm{CS})/N(\mathrm{CO})\la 10^{-4}$. 

The disk excitation temperatures of the HCO$^+$(1-0) and HCO$^+$ (3-2) transitions are shown in bottom of Fig.~\ref{ex} 
(left and right panels, respectively). The 3-2 rotational line is excited at about 10 times higher density than the critical 
density for HCO$^+$(1-0), $n_\mathrm{cr}\sim10^5$~cm$^{-3}$, leading to the lower excitation temperatures in the former 
case. Since HCO$^+$ ions are only $\sim10$ times more abundant than CS molecules, the HCO$^+$(1-0) and HCO$^+$(3-2) lines 
have a low optical depth of about $10^{-3}$. The disk excitation map for the C$^{18}$O(2-1) emission is similar to that of 
CO(2-1) and therefore is not shown.

Note that the CS(2-1) and HCO$^+$(1-0) excitation temperatures show a broad zone in the disk intermediate layer, where 
$T_\mathrm{ex}<0$ (light gray areas in the bottom left and top right panels, Fig.~\ref{ex}). Here a non-LTE effect plays a 
role. The inversion in the rotational level populations is caused by a specific ratio between collisional and radiative 
(de-) excitation probabilities. In those disk parts, where the concentration of CS and HCO$^+$ is rather low, 
their levels are excited and de-excited by collisions and only radiatively de-excited (but not excited). Thus, the LTE 
condition is broken and level populations do not follow the Boltzmann distribution. This effect for the CS molecule has 
been considered by \citet{CS-invers} in detail. However, such inversion does not lead to significant maser amplification 
of the line intensity even for the most favorable case of the edge-on disk. As soon as the optical depth of the line 
approaches unity (which is only possible in radial direction), stimulated radiative transitions become operative. They 
additionally excite and de-excite the level populations and destroy the inversion.

We do not present the resulting excitation temperatures for the envelope model because the excitation conditions 
there do not change as strongly as in the case of the disk, and resemble those of the low-density disk surface. 
Consequently, the CS(2-1) and HCO$^+$(3-2) emission lines are not thermalized anywhere in the envelope, 
$5~\mathrm{K} \la T_\mathrm{ex} \la 10~\mathrm{K}$, whereas HCO$^+$(1-0) is partly thermalized in the inner part at 
$r \la 800$~AU: $5~\mathrm{K} \la T_\mathrm{ex} \la 25~\mathrm{K}$. Finally, the CO(2-1) and C$^{18}$O(2-1) lines are 
thermalized in the entire envelope with $T_\mathrm{ex} \approx T_\mathrm{kin}\sim30$~K. 

A typical computation for a $56 \times 55$ grid model, 11 transitions, and the optically thick CO(2-1) line needs 
about 2 days on a Pentium IV 2.4~GHz PC. In the extremely optically thin approximation applied to simulate the radiative 
transfer in the CS(2-1), CS(5-4), C$^{18}$O(2-1), HCO$^+$(1-0), and HCO$^+$(3-2) lines, the photon ray tracing and ALI 
scheme are not used. In this case, the level populations are computed within a few seconds. The validity of this 
approach is verified by comparing once the level populations of the CS(2-1), C$^{18}$O(2-1), HCO$^+$(1-0), and 
HCO$^+$(3-2) transitions calculated with the full line radiative transfer and in the limit 
of the extremely low optical depth. It is found that both methods yield very similar 
results both for the disk and envelope model.

\section{Results of the line radiative transfer modeling}
\label{res}

In this section, we confront the theoretical model of the AB Aur system with the observational data and constrain basic 
parameters of the disk and envelope. It would be mathematically more correct to compare the interferometric data with the 
model results in the {\it uv}-plane in order to avoid the highly nonlinear deconvolution procedure \citep[see, e.g.][]{GD98},
but this is only meaningful for low-noise data, when model parameters can be determined with a high accuracy, e.g., 
by utilizing chi-square minimization. However, we follow a more illustrative way and face the synthesized HCO$^+$(1-0)
interferometric map {\em directly} with the observed line profiles. Another limiting factor for our numerical simulations is
the computational time. Typically, it takes about 10-30 minutes to generate one synthetic single-dish spectrum and $\sim 2$
days to synthesize the entire interferometric map with a 2.4~GHz Pentium IV machine. Together with chemical and non-LTE
line radiative transfer calculations, a total computational time for one modeling run can be as long as about 3 days. 
Therefore, a full $\chi^2$-minimization in the space of all parameters to be constrained is not feasible for our 
approach, and one has to estimate these values and their uncertainties with a more robust analysis. 

This analysis is based on the fact that many model parameters can be constrained {\em independently} from the others in a
subsequent way (``step-by-step''), starting from suitable initial guesses. First, we use the first-order estimates for the 
AB Aur disk orientation and determine more precisely the value of the disk inclination by fitting the width of the central 
HCO$^+$(1-0) spectrum in the interferometric map. Next, with this best-fit value we derive the disk positional angle by 
fitting the asymmetry of the observed line profiles out of the map center. After the best-fit disk orientation is found,
we investigate how the radial gradient of the normalized synthetic line intensities depends on the assumed disk radius and
choose accordingly the best-fit value of the disk size. Finally, the derived disk orientation and radius allow us to
constrain the total disk mass because this value defines the absolute intensities of the 
modeled HCO$^+$(1-0) lines on the interferometric map. The uncertainties of the best-fit disk parameters are 
found by the same iterative way.

Consequently, we constrain temperature, density structure, and mass of the inner, shielded part of the AB Aur envelope 
using the best-fit disk orientation and radius. First, we determine the temperature of the envelope and its uncertainty 
by modeling the intensity of the optically thick CO(2-1) line (see Sect.~\ref{env_model}). Second, with 
the best-fit envelope temperature, we iteratively find the value of the radial density gradient, initial density at 
the envelope inner radius, and their uncertainties by fitting the observed single-dish HCO$^+$(1-0), HCO$^+$(3-2), 
C$^{18}$O(2-1) and CS(2-1) spectra.

\subsection{Interferometric HCO$^+$(1-0) map}
\label{disk_res}
We present the results for our best-fit model of the AB Aur disk. Then we discuss how parameters of this model are actually
derived. This model is found after approximately 15 subsequent iterative steps of the modeling.

\begin{figure}
\includegraphics[width=0.87\textwidth, angle=270,clip=]{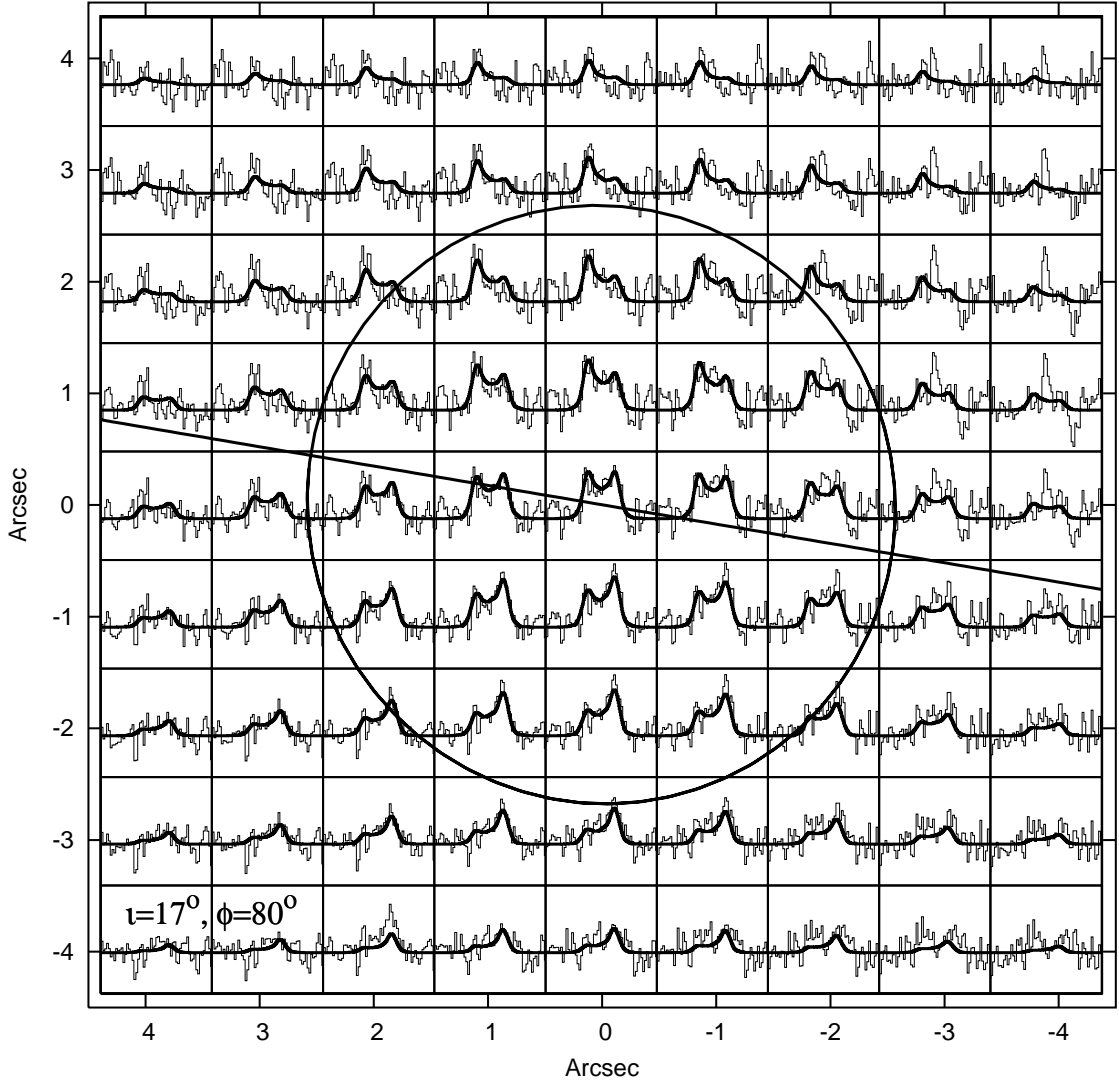}
\caption{Synthetic (thick line) and observed (thin line) HCO$^+$(1-0) interferometric spectra of 
the AB Aur disk are compared (these lines are marked red and blue in electronic edition). The
vertical (Dec) and horizontal (RA) axes are offsets in arcseconds from the disk center (standard
``N--E'' observational orientation). To produce every synthetic spectrum, we convolve it with the 
Gaussian beam of $5.87\arcsec$. The intensity of all spectra spans the same $[-0.12~\mathrm{K}, 0.195~\mathrm{K}]$
range expressed in units of the main beam temperature (see Table~\ref{lrt_par}). The size of the 
adopted disk model is depicted by solid circle, while the projection of the disk rotational axis 
on the sky plane is shown by straight line (both appear green in electronic edition). The best-fit values of the 
inclination ($\iota=17\degr$) and positional ($\phi=80\degr$) angles are indicated in the lower 
left corner of the plot.}
\label{map}
\end{figure}

The simulated and observed interferometric HCO$^+$(1-0) maps are compared in Fig.~\ref{map}. As can 
be clearly seen, intensities and widths of the synthetic spectra match well the observed values, 
$T_\mathrm{mb}\sim0.1$~K and $\Delta V_\mathrm{obs}\approx2.3$~km\,s$^{-1}$. Moreover, the synthesized line 
profiles in general follow the shape of the observed spectra on the entire interferometric map. This proofs again that the Keplerian law is a reasonable representation of the disk global velocity field.

The low intensities of the observed interferometric spectra, $T_\mathrm{mb}\sim0.1$~K, suggest that the HCO$^+$(1-0) line 
is optically thin. Given a typical excitation temperature in the disk for this transition, $T_\mathrm{ex}\sim100$~K (see
Fig.~\ref{ex}, lower left panel), the optical depth of HCO$^+$(1-0) can be roughly estimated as
$\tau \sim T_\mathrm{mb}/T_\mathrm{ex}\approx10^{-3}$, exactly the value we have found in Sect.~\ref{texc}.

The double-peaked shape of the synthetic line profiles and their asymmetry are the result of the Gaussian convolution 
with the $5.87\arcsec$ beam over the inclined disk of a similar size (see Table~\ref{disk_par}). Every 
HCO$^+$(1-0) spectrum is a beam-weighted average of the emission generated in various disk regions that have a broad 
velocity range (from negative to positive values in respect to the system velocity). The central spectrum on the 
interferometric map has a symmetric profile because disk locations with various velocities contribute equally to the
formation of this line. Similarly, all spectra located along the projection of the disk rotational axis on the sky plane 
(zero-velocity line, $V=V_\mathrm{lsr}$) are also symmetric (compare the line profiles nearby straight line and in 
perpendicular direction in Fig.~\ref{map}). All other spectra have asymmetric line profiles due to the beam weighting over 
the disk parts that have a lack of either blueshifted or redshifted emission components. The contrast between the left and 
right intensity peaks of the HCO$^+$(1-0) lines depends on the orientation of the beam in respect to the zero-velocity 
line and is maximal in orthogonal direction. 

The convolution with a beam size comparable to the size of the adopted disk model leads to the non-zero HCO$^+$(1-0) 
intensities beyond the disk outer boundary (depicted by circle in Fig.~\ref{map}). It is worth to mention here that the 
convolution with a beam size that is several times smaller than the disk size would produce a synthetic map of the narrow
single-peaked spectra with different velocity shifts in different disk parts, $|{\Delta V}| \propto r^{-0.5}$.

Note that there is a region in the interferometric map out of the disk ($x\sim-4\arcsec$, $y\sim2\arcsec$), where the 
observed HCO$^+$(1-0) lines have a peculiar shape that is far from the modeled one (upper right corner in Fig.~\ref{map}).
This shape has a $\sim-0.1$~K absorption feature with about $2$~km\,s$^{-1}$ offset from the system velocity 
$V_\mathrm{lsr}$ and a $\sim0.2$~K emission peak centered at $V_\mathrm{lsr}$, closely resembling an inverse P Cygni 
profile. It has been checked that the profiles of these ``emission-absorption'' spectra are stable to the data reduction 
with various CLEANing parameters and weightings. As we mentioned in Section~\ref{pdbi}, the detected 
HCO$^+$(1-0) continuum is low, which contradicts with the absolute intensity of the 
absorption peak. Thus, this is a ``ghost'' feature in the spectra, though the strength, width, and position of the 
central peak is consistent with the emission arising in low-velocity envelope regions. Moreover, the peculiarity in the 
observed profiles is only seen in a part of the full interferometric map.

The presence of a small dense structure in the disk outer region or nearby envelope could
be responsible for the appearance of such emission-absorption spectra in the deconvolved interferometric map. Why this 
structure has to be small and dense can be easily explained: its small ($\la2\arcsec$) angular extent is a necessary 
condition to produce the peculiar profiles in a confined area of the whole $12\arcsec\times13\arcsec$ spectral map only, 
while a high density contrast is needed for the interferometer to ``see'' the presence of such an inhomogeneity. 
Interestingly, \citet{Fea04} have resolved the AB Aur system at the $\sim100$~AU scale using the {\it Subaru Coronographic 
Imager and Adaptive Optics} at near-infrared wavelengths and discovered a few spiral arms and a knotty structure 
associated with a circular inhomogeneous structure of $\sim 580$~AU radius (see Fig.~2 and 3 therein). Thus, it is likely 
that one of these local and compact structures with enhanced density has been detected during our PdBI observational 
campaign.

\begin{figure}
\includegraphics[width=0.44\textwidth]{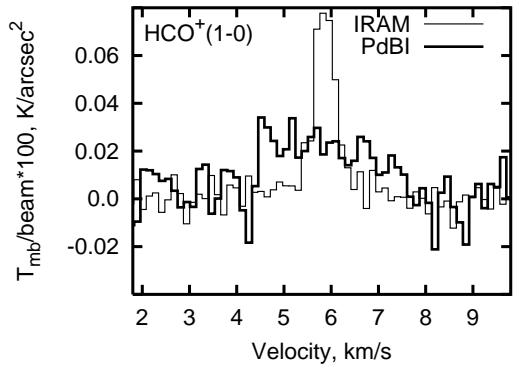}
\caption{Normalized single-dish IRAM spectrum (thin line) of the HCO$^+$(1-0) emission in comparison 
with the average HCO$^+$(1-0) interferometric spectrum (thick line) obtained with PdBI (the lines are 
blue and red in electronic edition). In the former case, the intensity of the spectrum has been scaled down 
by the square of the IRAM HCO$^+$(1-0) beam size of $29\arcsec$, while in the latter case the intensity 
has been averaged over a total of $12 \times 13$ ($1\arcsec \times 1\arcsec$) various interferometric HCO$^+$(1-0) PdBI 
spectra. 
}
\label{space}
\end{figure}

The fact that the interferometer probes only dense and compact matter is illustrated in Fig.~\ref{space}, where
we compare the observed beam-normalized HCO$^+$(1-0) IRAM 30-m and averaged PdBI spectra. The IRAM line profile is narrow, 
$\approx1$~km\,s$^{-1}$ and single-peaked, while the PdBI spectrum is about 4~km\,s$^{-1}$ wide and has a double-peaked 
shape. Furthermore, their normalized intensities differ by a factor of 2. The reason for such a diversity in the
observed spectra is that the IRAM 30-m antenna has $29\arcsec$ beam for the HCO$^+$(1-0) transition, covers large 
spatial scales, and therefore is not capable in detecting the emission from the small AB Aur disk due to the huge beam 
dilution. Instead, the HCO$^+$(1-0) emission from the surrounding envelope is only observed. Contrary, the $\approx5\arcsec$ 
PdBI beam is sensitive to small spatial scales and thus does not ``feel'' the emission that comes from the low-density and 
extended envelope (e.g., see the spectrum at the position $x=4\arcsec$, $y=-4\arcsec$ in Fig.~\ref{map}). 

\subsubsection{Disk orientation}
\begin{figure}
\plotone{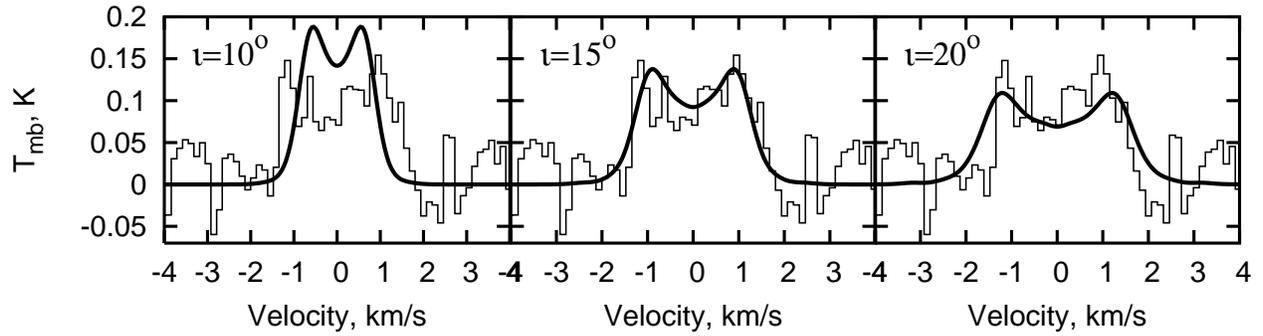}
\caption{Observed (thin line) and synthesized (thick line) HCO$^+$(1-0) spectra at the center of 
the interferometric map (these lines are red and blue in electronic 
edition, respectively). Three different cases are shown, namely, the inclination angle 
$10\degr$ (left panel), $15\degr$ (middle), and $20\degr$ (right panel). The observed line profile 
is consistent with a value of the inclination angle between $15\degr$ and $20\degr$.}
\label{incl}
\end{figure}

First, the disk inclination angle $\iota$ is determined by comparing the observed HCO$^+$(1-0) line profile at the center of 
the interferometric map with the corresponding synthetic spectrum. The modeled central line has a symmetric double-peaked 
profile independent of the actual value of the disk positional angle (see Fig.~\ref{map}, $x=0$, $y=0$). However, the line 
width and intensity do depend on the assumed value of the inclination angle, as shown in Fig.~\ref{incl}. It can be 
clearly seen that a $10\degr$ inclination of the disk results in a too narrow ($\approx1$~km\,s$^{-1}$) modeled spectrum 
compared with the observed $2.2$~km\,s$^{-1}$ line width (Fig.~\ref{incl}, left panel). On the other hand, the disk 
inclined by $\iota=20\degr$ produces a slightly broader spectrum than observed (Fig.~\ref{incl}, right panel). Thus, the 
disk inclination angle has the best-fit value somewhere between these two limits, $\iota\approx15\degr$ 
(middle panel in the Figure). We take $17\degr$ as the best-fit inclination angle of the AB Aur disk.
Note that a similar approach to constrain the disk inclination was recently presented by \citet{tw_hya}, who 
applied it for the TW Hya disk based on the SMA observations in the CO(2-1) and CO(3-2) lines.

The accuracy of this value is mainly determined by the uncertainty in stellar mass and by the radial gradient of the 
disk surface density, whereas other model parameters play a minor role. 

The width of the modeled central HCO$^+$(1-0) line for a fixed value of the disk inclination angle and density gradient
varies as $\sqrt{M_*}$ (Keplerian law). We consider two values of the stellar mass, $M_*=2.0M_{\sun}$ and $M_*=3.0M_{\sun}$, 
and find that for the low-mass limit of $2.0M_{\sun}$ the best-fit disk inclination is $19\degr$, while for the 
$3.0M_{\sun}$ star this value is about $14\degr.5$. In addition, we use two disk models with shallower surface density 
profiles, namely, $p=-3/2$ (minimum-mass solar nebula) and $p=0$ (uniform disk). In this case, the HCO$^+$ column densities
peak at larger radii than for the reference model with $p=-5/2$. However, we regain that such a modification does not lead 
to a significant spread in the derived best-fit inclination angle. For the uniform disk model the best-fit value is 
$\iota=23\degr$, whereas for the minimum-mass solar nebula it is $19\degr$. Taking into account all these uncertainty 
factors, we estimate that the AB Aur disk is inclined by $17^{+6}_{-3}\degr$.

\begin{figure}
\includegraphics[width=0.3682\textwidth,clip=]{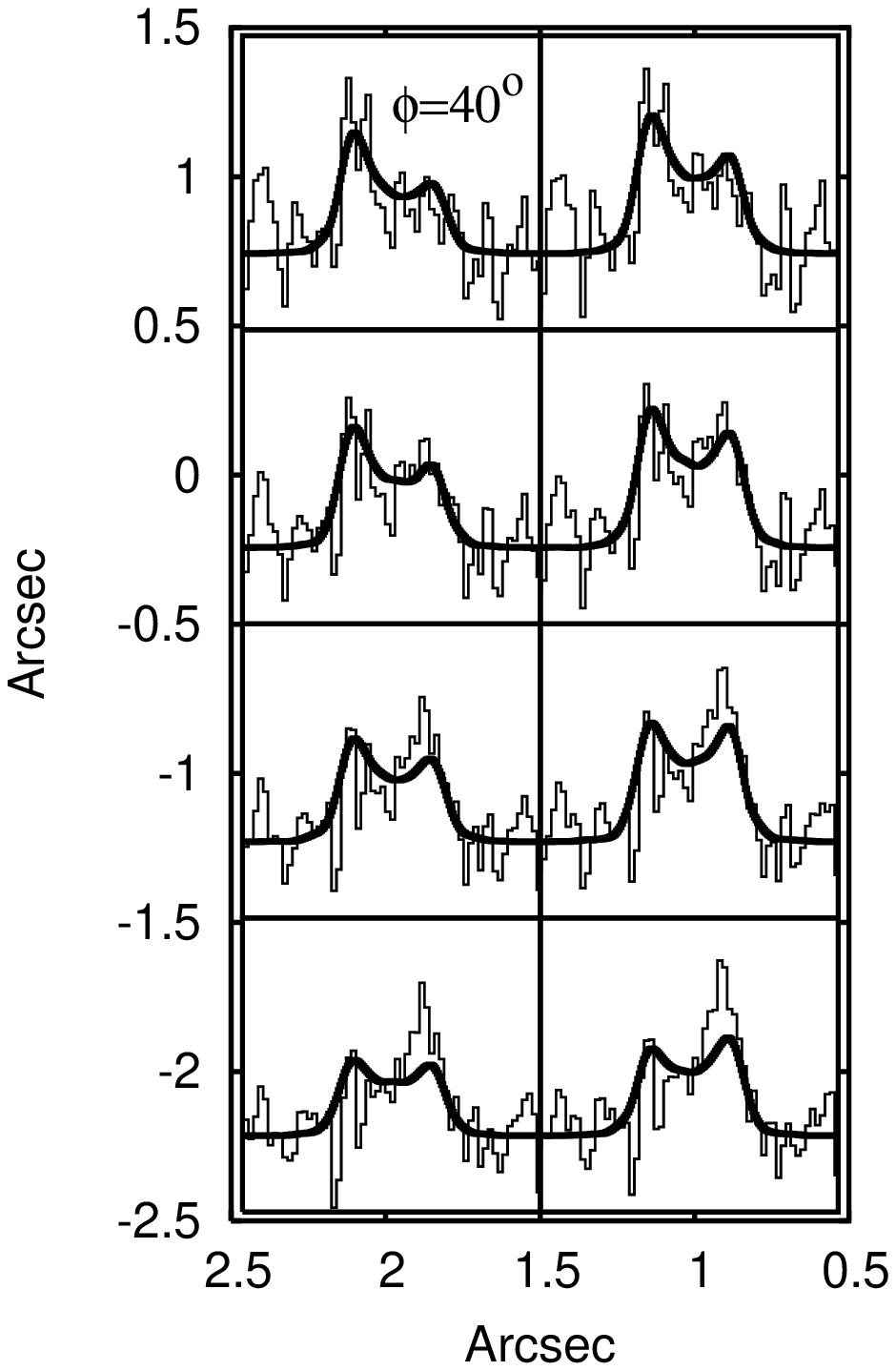}
\includegraphics[width=0.28\textwidth,clip=]{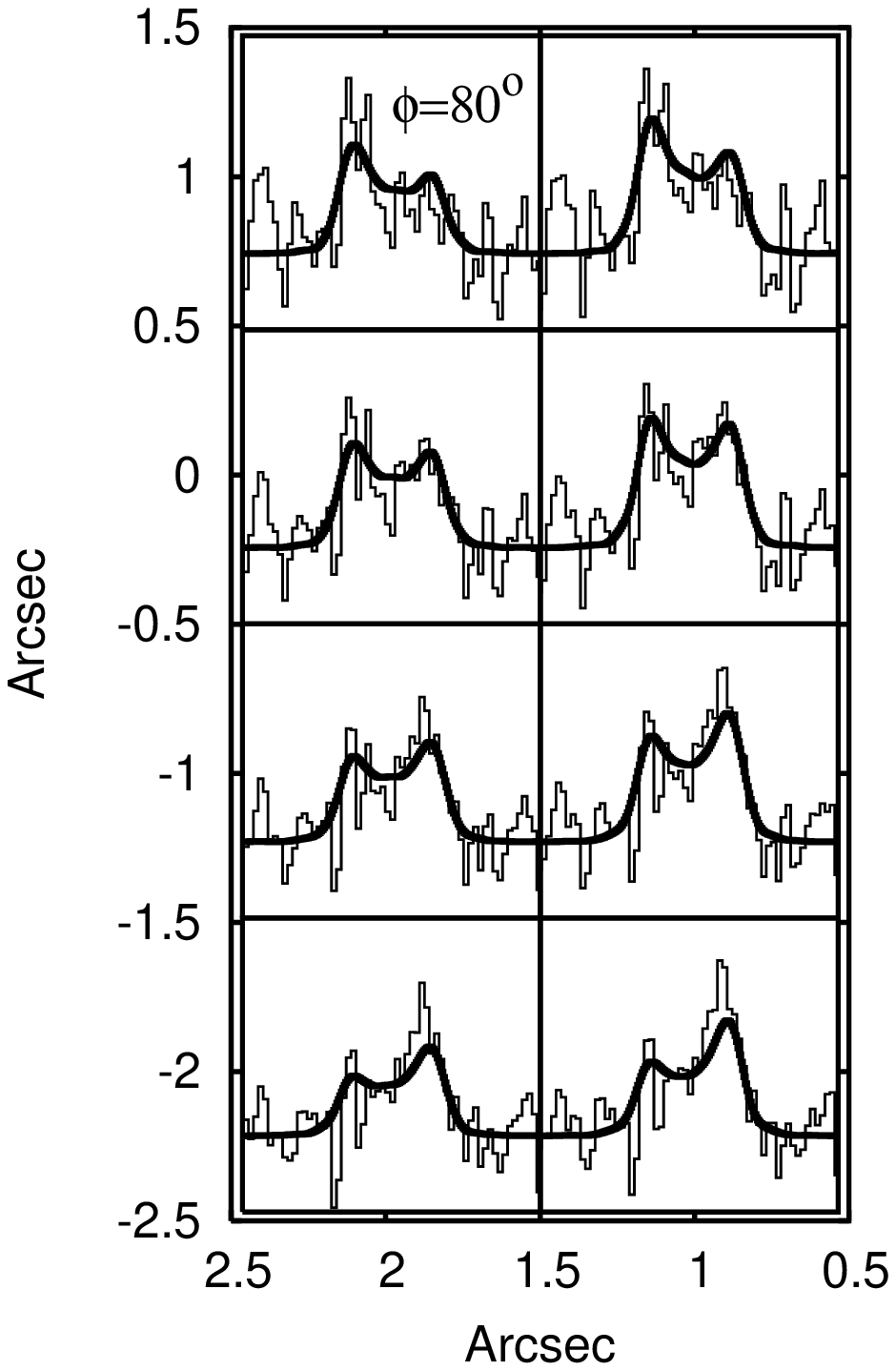}
\includegraphics[width=0.28\textwidth,clip=]{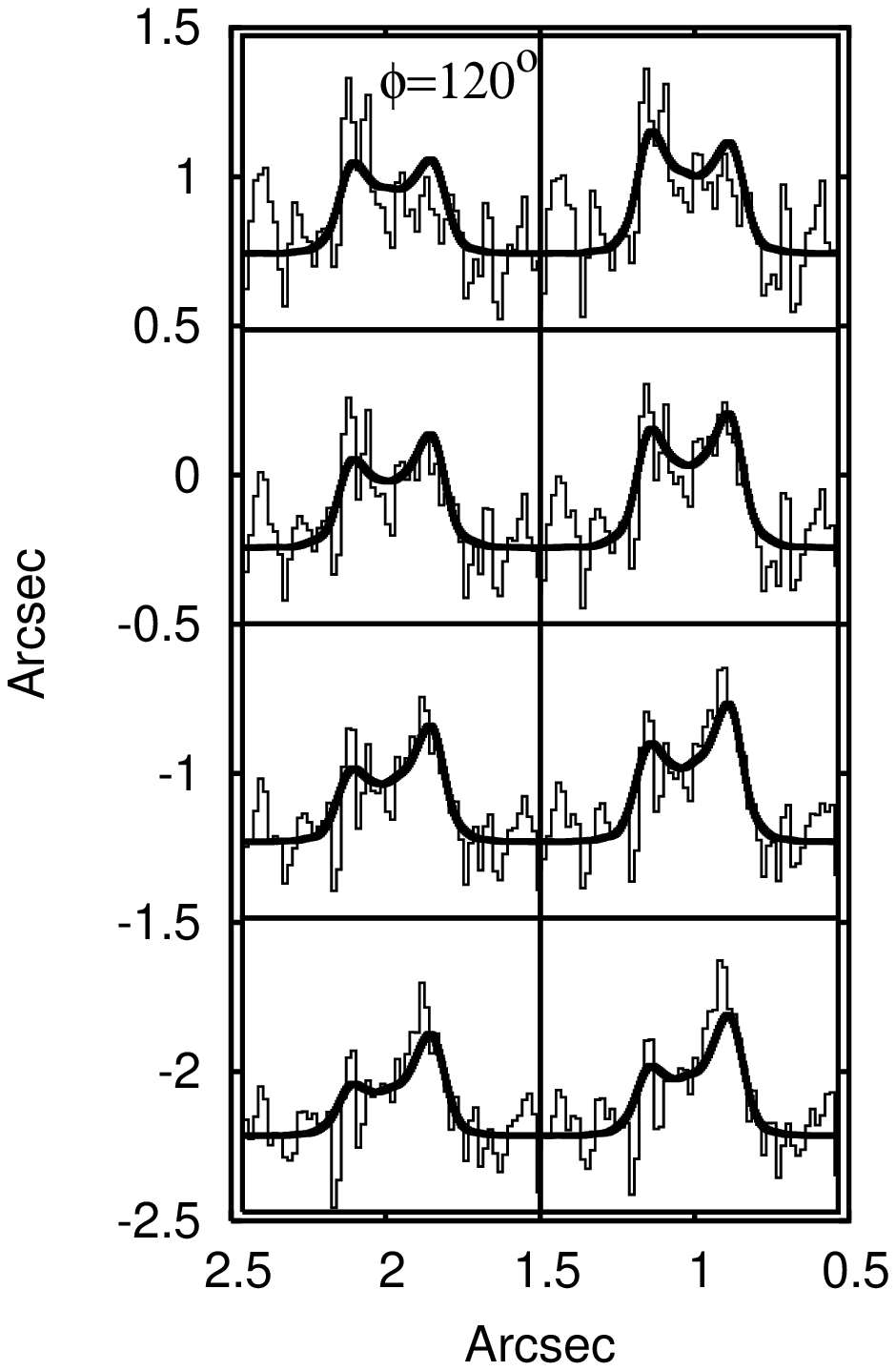}
\caption{Comparison between the observed (thin line) and synthetic (thick line) HCO$^+$(1-0) spectra for
a part of the whole interferometric map outside of the disk center (the lines are colored red and blue 
in electronic edition). The inclination angle is fixed, $\iota=17\degr$, whereas three different 
values of the positional angle are considered: $\phi=40\degr$ (left panel), $\phi=80\degr$ (middle),
and $\phi=120\degr$ (right panel). The overall comparison of the acquired and modeled line profiles
favors a value of the disk positional angle between $\phi\sim60\degr$ and $100\degr$. 
}
\label{posa}
\end{figure}

The value of the disk positional angle is constrained in a similar way. In Section~\ref{pdbi} a
first-order observational estimate, $\phi\sim90\degr$, has been obtained. In order to determine it better, we use the best-fit inclination angle and consider three different values of the positional angle: $\phi=40\degr$, 80$\degr$, and 120$\degr$. The resulting
modeled spectra are compared with the observational data at off-central positions of the interferometric map in 
Fig.~\ref{posa}. 

The most noteworthy changes in the synthetic HCO$^+$(1-0) lines are seen in two spectra at $x=2\arcsec$, $y=-2\arcsec$ 
(S1) and $x=2\arcsec$, $y=1\arcsec$ (S2). The disk positional angle $\phi=40\degr$ can be essentially ruled out 
because the asymmetry of the observed S1 profile is not properly fitted with this model (see the lower left corner of the left 
panel in Fig.~\ref{posa}). Contrary, with $\phi=120\degr$ it is not possible to explain the observed shape of the S2 
spectrum, see the left upper corner of the right panel in the Figure. Finally, the best fit to the asymmetry of the 
observed line profiles is obtained with the disk positional angle of $80\degr$ (middle panel in Fig.~\ref{posa}). However, 
this value cannot be constrained as accurately as the disk inclination angle since the noisy observational data can be 
fitted equally well with any other positional angle between $60\degr$ and $100\degr$. Therefore, the best-fit value of the 
disk positional angle is $\phi=80\degr\pm30\degr$ .

The derived orientation of the AB Aur disk, $\iota=17^{+6}_{-3}\degr$ and $\phi=80\pm30\degr$, is in reasonable
agreement with the recent high-resolution NIR observations of \citet{Eea03}, who have successfully 
reproduced the interferometric visibilities with uniform disk ($\iota=26^{+10}_{-19}\degr$, $\phi=128^{+30}_{-45}\degr$), 
accretion disk ($\iota=27^{+13}_{-17}\degr$, $\phi=105^{+34}_{-20}\degr$), and ring ($\iota=28^{+10}_{-18}\degr$, 
$\phi=144^{+17}_{-51}\degr$) models. It is also consistent with the disk inclination of $30\pm5\degr$
and positional angle $\phi=58\pm5\degr$ determined by \citet{Fea04} using isophoto fitting of the Subaru 
coronographic NIR image. Furthermore, the observational results of \citet{MGea99,MGea01} and \citet{Grea99}, as well as 
the measured low visual extinction toward this star \citep[e.g.,][]{Rea01}, favor to the face-on orientation of the AB Aur 
system ($\iota<45\degr$).

However, a nearly edge-on $76\degr$ inclination angle has been derived by \citet{MS97} from the analysis of the 
mid-resolution $\sim5\arcsec$ interferometric image of the AB Aur system obtained with the OVRO array in the $^{13}$CO(1-0) 
line. In many later studies aimed at the modeling of the AB Aur SED this value of the disk inclination 
has been adopted to constrain the model parameters \citep[see, e.g.,][]{Mea99,Dea03}, though \citet{Nea01} have used a 
more correct value of $30\degr$. 

\subsubsection{Disk radius and mass} 
With determined values of the AB Aur disk inclination and positional angles, we continue our step-by-step analysis and 
put constraints on the size and mass of this object.

There is no such controversy in the literature regarding the size of the AB Aur disk as in the case of  
orientation. The resolved disk radius has been determined as $\sim 400$-$600$~AU \citep{MS97,Fea04}.
We use the value $R^\mathrm{disk}_\mathrm{out}=400$~AU as an initial guess for the modeling \citep[see, however,][]{Nea01}.
A disk model of that size successfully reproduces the radial gradient of the observed HCO$^+$(1-0) intensities
(see Fig.~\ref{map}). In addition, we consider a smaller disk with $R^\mathrm{disk}_\mathrm{out}=200$~AU and find that this 
model still provides a reasonable fit to the radial decline of the observed line intensities, but
the width of the synthesized spectra is slightly broader than observed $\Delta V_\mathrm{obs}\approx2.3$~km\,s$^{-1}$.
To get the correct line widths, we adopt a smaller inclination angle of $15\degr$, which is close to the lower limit 
of the best-fit disk angle, $\iota=17\degr-3\degr=14\degr$. An even smaller $100$~AU
disk model shows a too rapid decrease of the calculated intensities with radius and hence is not in agreement with the data. 

We cannot determine an upper limit of the disk size by the same way because the gas in the outer disk regions is cold and 
diffuse. As shown in Sect.~\ref{texc}, the HCO$^+$(1-0) transition is hardly excited under these conditions; moreover, 
the HCO$^+$ abundances are low (see Fig.~\ref{colden}). As a result, the synthetic interferometric 
spectra almost do not change even if we take into account far-distant disk parts. Thus, as the upper limit we use the 
largest value of the resolved AB Aur disk, $R^\mathrm{disk}_\mathrm{out}=600$~AU.

Summarizing our findings, we estimate the best-fit size of the AB Aur circumstellar disk as 
$R^\mathrm{disk}_\mathrm{out}=400\pm200$~AU. 

The best-fit disk model with $R^\mathrm{disk}_\mathrm{out}=400$~AU and $\iota=17\degr$, which is shown in 
Fig.~\ref{map}, has a mass $M_\mathrm{disk}=1.3\cdot10^{-2}\,M_{\sun}$ (Table~\ref{disk_par}). To constrain the latter 
value, we rely on the fact that the absolute intensities of the calculated HCO$^+$(1-0) spectra are sensitive to the HCO$^+$ 
abundances and thus column densities in the disk due to the low optical thickness of this line:
\begin{equation}
\begin{array}{l}
T_\mathrm{mb} \propto \displaystyle \int_{\rm beam}<T_\mathrm{ex}(r)>N_{\mathrm{HCO}^+}(r)r\exp(-(\frac{r}{r_{\rm beam}})^2)dr \sim \\
\displaystyle <T_\mathrm{ex}(r_\mathrm{max})>N_{\mathrm{HCO}^+}(r_\mathrm{max})r_\mathrm{max}\exp(-(\frac{r_\mathrm{max}}{r_{\rm beam}})^2)\Delta r_\mathrm{max}.
\end{array}
\label{tmb}
\end{equation}
Here, $<T_\mathrm{ex}(r)>$ and $N_{\mathrm{HCO}^+}(r)$ are the averaged excitation temperature and HCO$^+$ column densities
at a certain radius $r$, respectively, while the term $r\exp(-(r/r_{\rm beam})^2)$ accounts for convolution 
with the Gaussian beam of the radius $r_{\rm beam}$. Using the column densities computed in Sect.~\ref{chem_res} and 
disk excitation temperatures of the 
HCO$^+$(1-0) line calculated in Sect.~\ref{texc}, we find that expression~(\ref{tmb}) has a global maximum at 
$r_\mathrm{max}\approx 120$~AU (with dispersion $\Delta r_\mathrm{max} \sim 60$~AU) {\em independent} of the considered disk 
model. Thus, the HCO$^+$ line profiles are mainly determined by the {\em local} emission generated in the disk regions at 
$r_\mathrm{max}$. Indeed, the width of the observed spectra can be calculated by the following equation:
\begin{equation}
\label{width}
\Delta V_\mathrm{obs}\approx2V_\mathrm{Kepl}(120~\mathrm{AU})\sin(\iota)=2\cdot4.2~\mathrm{km}\,\mathrm{s}^{-1}
\cdot\sin(17\degr)=2.47~\mathrm{km}\,\mathrm{s}^{-1},
\end{equation}
which is very close to the actual value of $\approx2.3$~km\,s$^{-1}$. 

We find that for 10 times more/less massive disks the modeled HCO$^+$(1-0) interferometric lines are 7/4.5 times more/less 
intense than in the case of the best-fit model with $M_\mathrm{disk}=1.3\cdot10^{-2}\,M_{\sun}$ (reference model). At first
glance, it seems that the value of the disk mass can be accurately determined since the intensity of the resulting
spectra sensitively depends on this parameter. However, the observed HCO$^+$(1-0) flux suffers from the
calibration errors, which are $\ga10\%$. Furthermore, the distance toward AB Aur is derived with $\sim15\%$ uncertainty 
(see Table~\ref{star_par}), and hence the flux is uncertain by additional $30\%$. Thus, the intrinsic uncertainty of the 
observed line intensities is $\ga40\%$. Consequently, it results in a factor of $\sim3$ uncertainty in the best-fit value
of the disk mass. Nonetheless, the spread in $M_\mathrm{disk}$ is mostly defined by the uncertainties in those parameters 
of the model that strongly affect the resulting abundances of HCO$^+$ (see discussion in Sect.~\ref{chem_res}) and thus 
the intensity of the synthesized spectra (expression~\ref{tmb}). 

Above all, the disk mass depends on the assumed gas-to-dust ratio because this parameter regulates the total amount of 
HCO$^+$ in the disk at large radii, $r\ga100$~AU (see Fig.~\ref{colden}, left panel). It is found that for the ratio 
$m_\mathrm{gd}=1000$ the simulated HCO$^+$(1-0) interferometric lines have intensities $T_\mathrm{mb}\sim 0.047$~K, which 
is 2.1 times lower than observed. Contrary, the model with low gas-to-dust ratio, $m_\mathrm{gd}=10$, produces 4 times more 
intense lines than the observed spectra with $T_\mathrm{mb}\sim 0.1$-0.2~K. Thus, such unrealistically high variation of 
the gas-to-dust ratio introduces a factor of 5 uncertainty in the best-fit disk mass 
$M_\mathrm{disk}=1.3\cdot10^{-2}\,M_{\sun}$. Moreover, we realize that it also affects the disk inclination angle in a 
sense that the disk model with $m_\mathrm{gd}=10$ requires $\iota=19\degr$ to fit the widths of the observed HCO$^+$(1-0) 
line profiles, while for the case of $m_\mathrm{gd}=1000$ it is $15\degr$, which is still within the proposed range of the 
disk inclination angles, $\iota\in[14\degr,23\degr]$. Similarly to the case of high gas-to-dust ratio $m_\mathrm{gd}=1000$,
the disk model with large $1\mu$m grains shows $\approx2$ times less intense spectra than the intensity of the acquired 
HCO$^+$(1-0) lines. To compensate for this decrease of the modeled line intensity, the disk mass has to be increased by a 
factor of 5 in respect to the reference value.

The next most important parameter determining the disk mass is the factor of $\sim2$ uncertainty of the computed HCO$^+$ 
abundances. In this case, the intensities of the synthetic lines vary by a factor of $\sim2.4$-3.5 compared with the 
observed values $T_\mathrm{mb}\la 0.2$~K. Therefore, for the model with 2 times increased abundances of HCO$^+$ the 
corresponding disk mass is only $20\%$ of the standard value $M_\mathrm{disk}=1.3\cdot10^{-2}\,M_{\sun}$, whereas for the 
case of the 2 times lowered HCO$^+$ abundances the disk mass is 1.8 times higher than for the reference model.

The parameters of the disk model that influence the evaluation of the disk mass to a smaller extent are the size and density
distribution. The uniform disk model with the radial gradient of the surface density $p=0$ and mass 
$1.3\cdot10^{-2}\,M_{\sun}$ results in 1.2 times higher intensity of the modeled spectra in comparison with the observed
line profiles. Consequently, the relevant disk mass constitutes $70\%$ of the mass of the standard model. In contrast, the 
model with the reduced outer radius $R^\mathrm{disk}_\mathrm{out}=200$~AU produces synthetic lines of 1.5 times lower 
intensity than the observed value of $\la0.2$~K. In this case, the best-fit mass of the disk is $50\%$ higher than 
the reference value of $1.3\cdot10^{-2}\,M_{\sun}$.

All considered model configurations and corresponding estimates of the AB Aur disk mass are summarized in 
Table~\ref{disk_mass}.

Overall, we constrain the AB Aur disk mass to $M_\mathrm{disk}\sim1.3\cdot10^{-2}M_{\sun}$ with a factor of 7 uncertainty. 
This best-fit value is in agreement with $M_\mathrm{disk}=2.1\pm0.9\cdot10^{-2}M_{\sun}$ determined by \citet{Tea01} 
from the 1.3mm flux, assuming dust opacities $\kappa_{1.3\mathrm{mm}}=0.01$~cm$^2$\,g$^{-1}$ and a gas-to-dust ratio of 100
\citep[see also][]{MS97}.

\subsection{The single-dish data}

\subsubsection{Temperature of the envelope}
\label{tenv}
\begin{figure}
\plotone{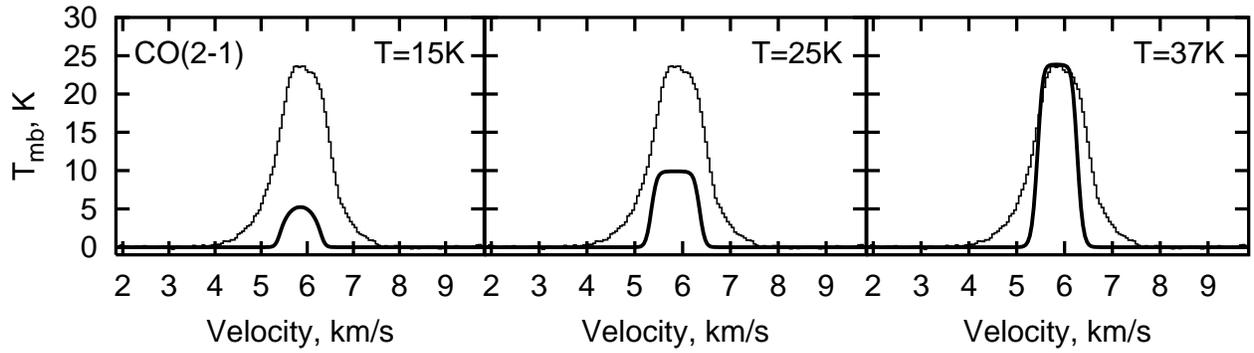}
\caption{Observed (thin line) and synthetic (thick line) single-dish CO(2-1) spectra are compared 
for three different envelope models (in electronic edition the lines are blue and red, 
respectively). From left to the right panel, the models with a fixed density distribution but 
different temperature of 15~K, 25~K, and 37~K are presented. As can be clearly seen, the
observed line intensity is reproduced only with the latter model. The high intensity 
($T_{\rm mb} \approx 25$~K) of the CO(2-1) spectrum suggests that this emission line is optically thick.}
\label{env_temp}
\end{figure}

As we have shown in Sect.~\ref{texc}, the observed single-dish CO(2-1) line is optically thick and thermalized, 
$T_\mathrm{ex}\sim T_\mathrm{kin}$. Therefore, the intensity of the CO(2-1) synthetic spectrum depends on the assumed value 
of the envelope temperature and does not depend much on the adopted density structure. We use this fact and consider a grid 
of the envelope models with various kinetic temperatures within the proposed initial range of $T=20$-40~K. In 
Fig.~\ref{env_temp}, we show the synthetic CO(2-1) line profiles calculated for three different envelope temperatures, 
namely, $T_{\mathrm{kin}}=15$~K (left panel), $T_{\mathrm{kin}}=25$~K (middle panel), and $T_{\mathrm{kin}}=37$~K (right 
panel). As can be clearly seen, the observed and modeled CO(2-1) line intensities are nearly the same only in the latter 
case, whereas for temperatures $T\la35$~K the computed line intensities are too low. Therefore, the value of 35~K is used 
in further modeling. The $\sim40\%$ uncertainty of the best-fit envelope temperature comes from the calibration error of the 
observed CO(2-1) flux and the distance uncertainty. Finally, we constrain the best-fit temperature of the inner AB Aur envelope 
to $T_\mathrm{env}=35\pm14$~K.

\begin{figure}
\plotone{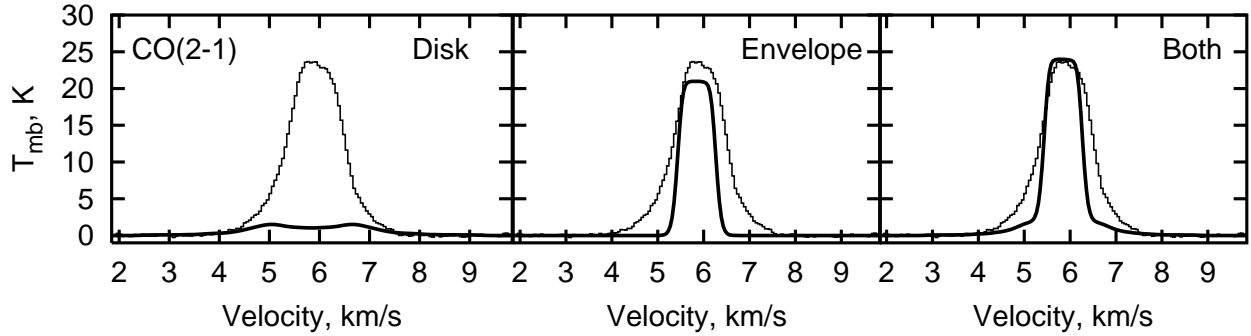}
\caption{Same as in Fig.~\ref{env_temp} but for three different models. In the left panel,
we present the case where only the disk model has been used in the line radiative transfer calculations, whereas in the
middle panel the result for the envelope model is shown. Finally, the combination of these two cases (disk-in-envelope
model) is shown in the right panel.}
\label{co2-1}
\end{figure}

In Fig.~\ref{co2-1}, we investigate what the relative contributions of the CO(2-1) emission generated by the disk and 
envelope on the resulting line profile are. The modeled disk emission shows very low intensity 
$T_\mathrm{mb}\approx 2~\mathrm{K}$ 
compared with the observed value of $24$~K, though the width of the synthetic spectrum is consistent with the observed 
$\approx2$~km\,s$^{-1}$ width (left panel). In contrast, the CO(2-1) emission from the envelope has a nearly correct 
intensity of $21$~K, but a too narrow width of 1~km\,s$^{-1}$ due to low infall velocities adopted in the model, 
$V(r) \sim 0.1$~km\,s$^{-1}$ (see middle panel in the Figure). The combination of 
both these models, the so-called ``disk-in-envelope'' model, results in the synthesized CO(2-1) profile with the correct 
intensity but still too narrow width at signal levels $T_\mathrm{mb}\sim5$-10~K (right panel in Fig.~\ref{co2-1}). 
As discussed in Sect.~\ref{iram}, emission arising in moving gas clouds along the line of sight to AB Aur may 
contaminate the observed CO(2-1) line profile. The chemical stability of the CO molecules to dissociative UV
radiation and low critical density to excite the 2-1 rotational transition support this suggestion. Recently,
\citet{Rea01} have studied with HST and FUSE the properties of H$_2$ and CO gases toward AB Aur. They have 
estimated a CO column density of $N_\mathrm{CO}=7.1\pm0.5\cdot10^{13}$~cm$^{-2}$, and found that the value of 
the CO velocity is consistent with the velocity of the star. Thus, this gas is indeed most likely associated with the 
nearby remnant envelope.

\subsubsection{Density structure of the envelope}
\label{env_den}
Using the best-fit envelope temperature $T_\mathrm{env}=35$~K, we simulate the chemical evolution and line 
radiative transfer for a grid of the envelope and disk-in-envelope models with various initial densities $\rho_0$ and 
density profiles within the range $p\in[-2,0]$. Then, the mass of the shadowed part of the AB Aur inner envelope can be 
calculated:
\begin{equation}
\label{env_mass}
M^\mathrm{sh}_\mathrm{env}=4\pi\frac{\theta}{90\degr}\rho_0r_0^3\frac{((r_1/r_0)^{p+3}-1)}{p+3},
\end{equation}
where $\theta=25\degr$ is the shadowing angle of the envelope (see Table~\ref{env_par}), and $r_0$ and $r_1$ are the inner
and outer envelope radii, respectively.

First, the intensity of the synthetic HCO$^+$(1-0) line is used as a criterion to determine whether the current guess of 
the initial envelope density $\rho_0$ (and thus the envelope mass $M^\mathrm{sh}_\mathrm{env}$) is appropriate or not 
because the observed HCO$^+$(1-0) emission almost entirely comes from the AB Aur envelope (see discussion in 
Sect.~\ref{texc}). The low optical thickness of HCO$^+$(1-0) implies that the corresponding calculated intensity is 
related to the HCO$^+$ abundances and can be easily scaled up and down by adjusting the mass of the envelope model until 
the observed intensity $T_\mathrm{mb}=0.55$~K is reached.

Second, with the updated envelope model that fits the observed HCO$^+$(1-0) line intensity, we verify whether the radial 
density gradient of this model is appropriate or not by comparing the observed and modeled HCO$^+$(3-2) to HCO$^+$(1-0) 
intensity ratios as well as the line profiles of C$^{18}$O(2-1), CS(2-1), and CS(5-4). The radial gradient of the infall 
velocity is calculated from the value of the density gradient as $-2-p$, while the initial infall velocity is determined
from the observed line widths, $V_{\mathrm{env}}(r_0) \sim 0.2$~km\,s$^{-1}$. 

Iterating these two steps of the modeling about 15 times, the final best-fit model of the AB Aur inner envelope is obtained
with the radial density profile $p=-1$ and thus infall velocity law 
$V_\mathrm{env}(r)=0.2\cdot(r/400\mathrm{AU})^{-1}$~km\,s$^{-1}$, initial density $\rho_0=9.4\cdot10^{-19}$~g\,cm$^{-3}$
($n_0 \approx 3.9\cdot10^5$~cm$^{-3}$), and mass $M^\mathrm{sh}_\mathrm{env}=4\cdot10^{-3}M_{\sun}$ (see 
Table~\ref{env_par}). 

\begin{figure}
\includegraphics[width=0.95\textwidth,clip=]{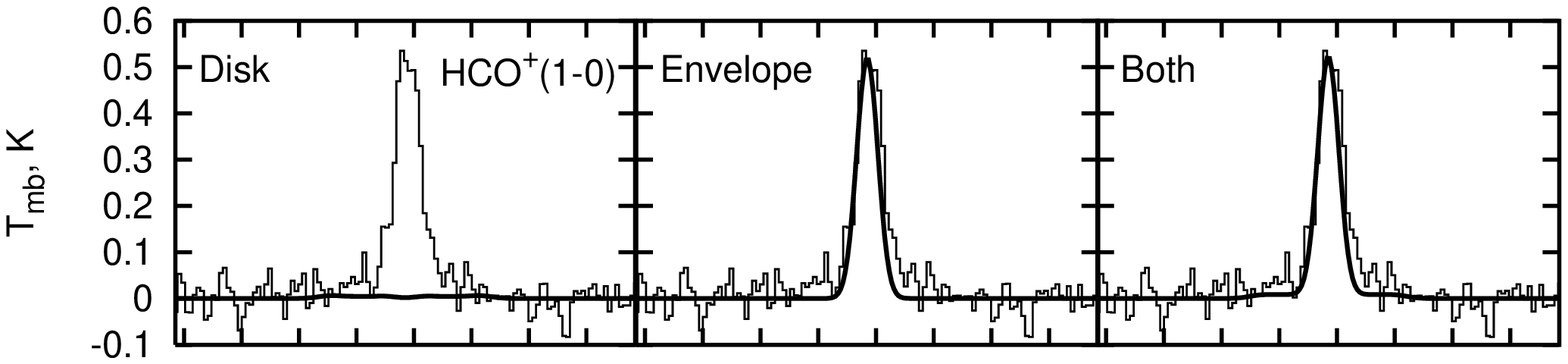}
\includegraphics[width=0.95\textwidth,clip=]{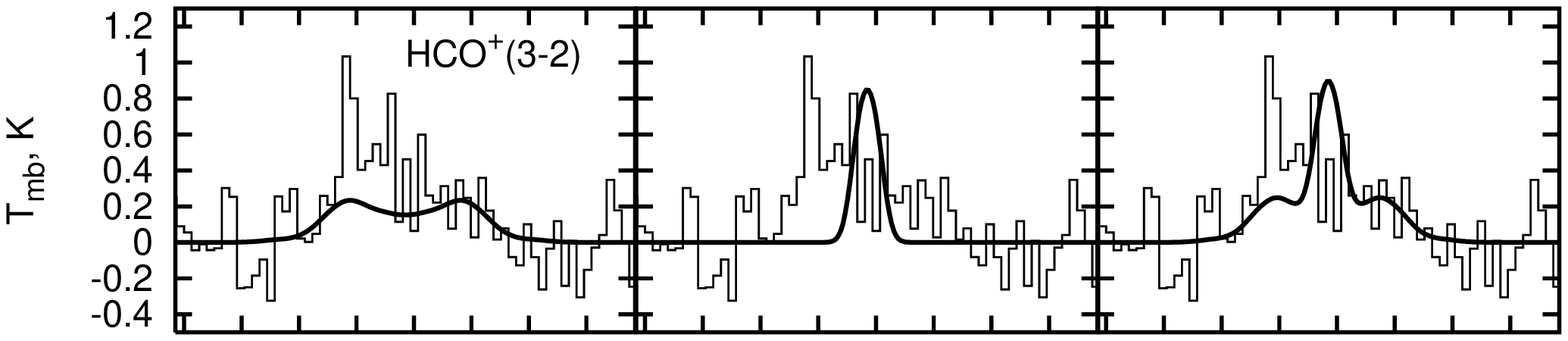}
\includegraphics[width=0.95\textwidth,clip=]{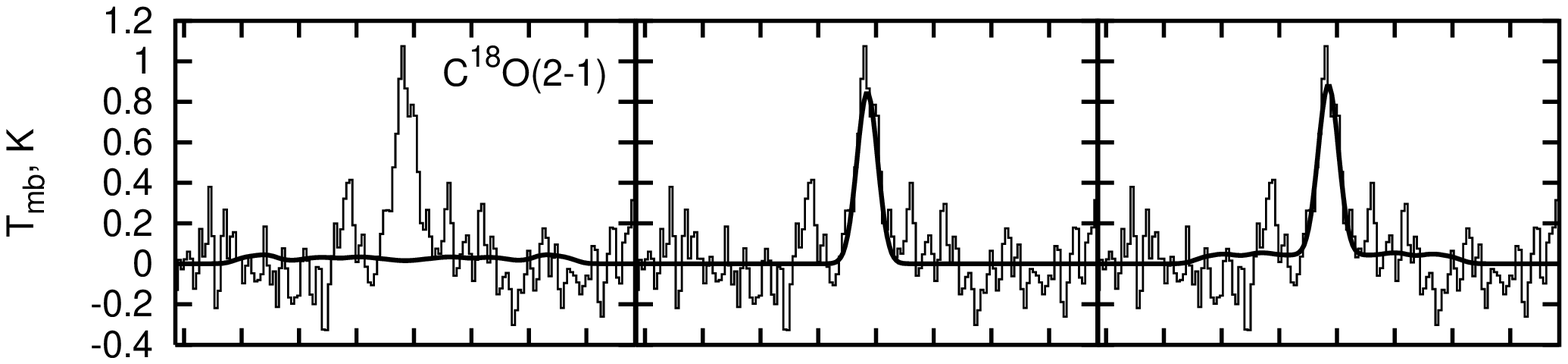}
\includegraphics[width=0.95\textwidth,clip=]{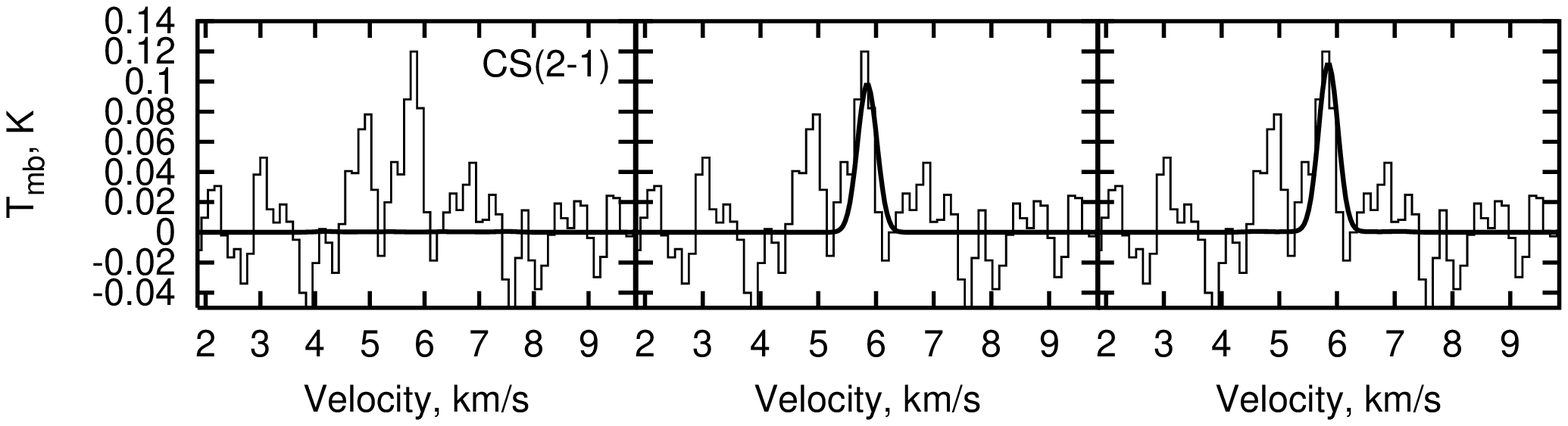}
\caption{Same as in Fig.~\ref{co2-1} but for the HCO$^+$(1-0) line (top row). In contrast with CO(2-1), 
this line is optically thin. This is also true for the HCO$^+$(3-2) spectrum, which is shown in the
second row from top. Note that the HCO$^+$(3-2) emission generated in the disk contributes much more to 
the resulting line profile than in the case of the HCO$^+$(1-0) line. The optically thin C$^{18}$O(2-1) and CS(2-1) 
lines are shown in the third and fourth rows from top, respectively. Overall agreement between the
observed and modeled spectra is only achieved with the disk-in-envelope model (right panels).}
\label{other_lines}
\end{figure}

The corresponding synthetic spectra are compared with the observed line profiles in Fig.~\ref{other_lines}. As can be 
clearly seen, intensities and widths of the synthesized and observed HCO$^+$(1-0), C$^{18}$O(2-1), and CS(2-1) lines are 
perfectly matched in the case of the envelope (middle panels) and disk-in-envelope (right panels) models, whereas the
contribution from the disk in the resulting spectra is negligible (left panels). In contrast, the observed HCO$^+$(3-2) 
emission comes partly from the AB Aur disk and partly from the surrounding envelope, therefore a reasonable fit to this 
line is only produced by the full disk-in-envelope model (see second row of plots in the Figure). Particularly, the broad
$\sim 2$~km\,s$^{-1}$ width of the observed HCO$^+$(3-2) spectrum is not possible to explain without accounting for the
emission from the AB Aur disk. Still, the synthetic HCO$^+$(3-2) spectrum peaks at about 1~km\,s$^{-1}$ higher 
velocity than the observed emission from the AB Aur envelope, which cannot be explained in the framework of the applied 
model.

The uncertainties of the derived parameters are estimated as follows. We adopt a factor of 7 uncertainties for the best-fit
value of the envelope mass and initial density at the inner envelope edge based on the results of a similar investigation 
performed for the AB Aur disk model (see extensive discussion in the previous section). The uncertainty of the determined
radial density profile $p=-1.0$ is found to be $\pm0.3$. This is because an envelope model with $p=0$ show $\sim2$ and 
3 times lower intensities of the synthetic C$^{18}$O(2-1) and CS(2-1) lines than the observed values of $1.1$~K and 0.11~K, 
respectively. For a model with $p=-1.5$, the intensity of the calculated HCO$^+$(3-2) spectrum is about 1.5 times higher 
than the observed value $T_\mathrm{mb}\approx0.9$~K. 

Note that in framework of the adopted model of the AB Aur system it is not possible to fit the observed CS(5-4) line profile 
since the intensity of the synthesized CS(5-4) spectrum is essentially zero. This line is excited at high densities, 
$n_\mathrm{cr}\ga10^6$-$10^7$~cm$^{-3}$, which are reached only in inner dense disk regions at $r\la150$~AU (see 
Fig.~\ref{disk}, right panel). Therefore, the intensity of the modeled CS(5-4) line becomes very low after the convolution 
with the $9\arcsec$ ($\sim 1000$~AU) IRAM beam. Consequently, a necessary requirement for a model of the AB Aur system to 
reproduce the observed CS(5-4) line intensity would be the presence of high-density media (clumps?) at $\ga100$-200~AU 
distance from the central star. 

We find that the intensity, width, and profile of CS(5-4) can be reproduced 
very well with a 5 times more massive model of the AB Aur envelope than the
best-fit one. However, in this case all other modeled single-dish lines are 
differ substantially from the observed spectra. Therefore, we try to match all single-dish 
data {\em simultaneously} with a clumpy envelope model. 
Surprisingly, it is possible only in one case: clumps that homogeneously fill 
$3\%$ of the entire $10\arcsec$ IRAM CS(5-4) beam, and which are about 150 times denser
than surrounding medium, $n \sim 5\cdot 10^7$~cm$^{-3}$. The total mass of these 
clumps is approximately equal to the mass of the best-fit envelope model, 
$M_{\rm clumps} \approx 4\cdot 10^{-3}M_{\sun}$, and consequently the mass of the clumpy 
envelope model is twice that of the reference best-fit model. The applicability of 
this clumpy envelope model cannot be proved with the adopted 2D approach and requires 
full 3D treatment, which is beyond the scope of the present paper.

The density distribution and mass of the best-fit model of the AB Aur envelope are consistent with the values derived by 
\citet{Eea04} from the modeling of the SED: $p=-1.4$, $\rho_0\approx1.6\cdot 10^{-18}$~g\,cm$^{-3}$ at $r=400$~AU, and 
$M_\mathrm{env}$ within $r\in[400~\mathrm{AU},2\,200~\mathrm{AU}]$ of about $2\cdot 10^{-2}M_{\sun}$. There is almost no 
difference between initial densities for both models, the radial density profiles are also similar, but the envelope masses
differ by a factor of 5. This is because we focus on the shadowed part of the inner envelope part that contains only 
$25\degr/90\degr \approx 28\%$ of the volume of the full sphere. If one takes into account $28\%$ of the total envelope 
mass determined by Elia et al., the resulting value is $5.5\cdot10^{-3}M_{\sun}$, which is close to the 
$4\cdot10^{-3}M_{\sun}$ mass of our best-fit model. 

Similarly, \citet{Fea02} have estimated the mass of the AB Aur envelope within 0.08~pc as $1M_{\sun}$ and obtained  a density profile 
of $-2<p<-1$. These values can be translated to the initial density at $r=400$~AU and mass of the shielded inner part of 
the envelope: $\rho_0\approx1.3\cdot 10^{-18}$~g\,cm$^{-3}$ and $M^\mathrm{sh}_\mathrm{env}\approx6.8\cdot 10^{-3}M_{\sun}$
for $p=-1$ and $\rho_0\approx3.7\cdot 10^{-17}$~g\,cm$^{-3}$ and $M^\mathrm{sh}_\mathrm{env}\approx5.4\cdot 10^{-2}M_{\sun}$
for $p=-2$, respectively. The mass and initial density of our best-fit envelope model 
($\rho_0=9.4\cdot10^{-19}$~g\,cm$^{-3}$ and $M^\mathrm{sh}_\mathrm{env}=4\cdot 10^{-3}M_{\sun}$) are close to the values of
Fuente et al. for the case of $p=-1$, while the steeper density gradient $p=-2$ requires about a 10 times more massive
envelope model to match the AB Aur SED.  

Finally, \citet{Mea99} have found by the modeling of the AB Aur SED with a disk-in-envelope model the total mass of the
envelope to be $M_\mathrm{env}\sim 0.03M_{\sun}$. Again, it can be translated to the value of $\sim 5\cdot10^{-4}M_{\sun}$ for 
the inner shadowed part, which is $\sim10$ times lower than in our case. Such a low value of the envelope mass is due to 
the uniform density distribution ($p=0$) at $r\ga120$~AU they have used in the calculations. 

Since we focus on the shadowed and inner region of the AB Aur envelope, it is difficult to calculate accurately the total 
mass of the envelope with our model. The mass of two unshielded lobes can be roughly estimated from the value of the
observed extinction toward AB Aur, $A_\mathrm{V}\la0.5$~mag. For our best-fit model of the shadowed part, the visual 
extinction is maximal in radial direction, $A^\mathrm{sh}_\mathrm{V}=2.5$~mag, which is about 5 times higher than for the 
unshielded region. If one assumes that this difference stems entirely from the density contrast between the shadowed and 
unshielded envelope parts and their radial density profiles and dust grain properties are the same, than the total 
envelope mass between 400~AU and 2\,200~AU is 
\begin{equation}
\label{env_mass_total}
M_\mathrm{env}\sim(1+\frac{A_\mathrm{V}}{A^\mathrm{sh}_\mathrm{V}}\,\frac{90\degr-\theta}{\theta})\,M^\mathrm{sh}_\mathrm{env},
\end{equation}
where $\theta=25\degr$ is the shadowing angle, $A_\mathrm{V}\la0.5$~mag, $A^\mathrm{sh}_\mathrm{V}=2.5$~mag, and
$M^\mathrm{sh}_\mathrm{env}=4\cdot 10^{-3}M_{\sun}$. According to this Equation, the total mass of the AB Aur inner 
envelope for the model discussed here is $M_\mathrm{env}\sim6\cdot 10^{-3}M_{\sun}$. Moreover, assuming that the best-fit density 
profile of the inner envelope model is also appropriate at larger distances
until the cloud outer border at 35\,000~AU, we calculate the mass of the {\em entire} 
AB Aur envelope, $M_{\rm env} \approx 1M_{\sun}$, which is exactly the value
measured by \citet{Fea02}.

The parameters of the density structure and mass of the envelope models considered above are summarized in 
Table~\ref{env_com}.

\subsection{Evolutionary status of the AB Aur system}
\label{evolution}

In this Section, we discuss the evolutionary nature of the 
AB Aur system, using the results of our modeling. 

With the reconstructed best-fit velocity profile of the AB Aur envelope (see Table~\ref{env_par}), 
one can estimate {\em independently} the mass accretion rate in this system: 
\begin{equation}
\dot{M}_{\rm acc}=3.16\cdot 10^7\,\rho_0\,V_{\rm env}(r_0)\,4\pi\frac{\theta}{90\degr}r_0^2,
\end{equation}
where $\rho_0=9.4\cdot10^{-19}$~g\,cm$^{-3}$ is the initial density at the disk outer edge 
$r_0=400$~AU and $\theta=25\degr$ is the shadowing angle. According to this expression, 
the mass accretion rate from the AB Aur envelope onto the disk is 
$\dot{M}_{\rm acc}\approx 4\cdot 10^{-8}M_{\sun}$~yr$^{-1}$. This value is very 
close to the measured mass accretion to the central star, 
$\dot{M}_{\rm acc}\sim 10^{-8}M_{\sun}$~yr$^{-1}$ \citep[see, e.g.,][]{Gea96}. 

If one assumes that the mass accretion rate is due to the viscous evolution of the disk only, then dispersal 
timescale for the $\sim 10^{-2}M_{\sun}$ AB Aur disk is $\tau_{\rm disk} \sim M_{\rm disk}/\dot{M}_{\rm acc} 
\sim 0.3$~Myr, which is too short in comparison with the $\sim 4$~Myr age of AB Aur. Thus, accretion from 
the envelope should play a major role in the evolution of the AB Aur system as a whole. The lifetime 
of the entire $\sim 35\,000$~AU envelope can be roughly estimated from the same principles, 
$\tau_{\rm env} \sim M_{\rm env}/\dot{M}_{\rm acc}\approx 1M_{\sun}/4\cdot 10^{-8}M_{\sun}$~yr$^{-1}$, 
which gives evolutionary timescale of about $25$~Myr. Therefore, we conclude that the AB Aur system will remain a 
Class~II object for the next few million years.

The estimated timescale for dispersal of the AB Aur envelope is in sharp contrast to the free-fall 
time from the envelope outer edge at 35\,000 AU on the central star, 
$\tau_{\rm ff}=\sqrt{3\pi/32G\rho} \approx 0.3$~Myr. Thus, contraction of the envelope is not free-fall 
and regulated by additional force(s) acting against gravitation, like thermal pressure, rotation, turbulence,
or magnetic field. 

The suggestion that the AB Aur envelope fully rotates can be essentially ruled out
given the fact that the widths of the observed single-dish lines are narrow, $\la 1$~km\,s$^{-1}$. 
Indeed, it is found that the model with only $20\%$ of the Keplerian rotation representing the regular 
velocity field in the envelope produces wider synthetic single-dish lines than observed. On the other
hand, conservation and redistribution of initial angular momentum of the natal cloud out of which the 
AB Aur system has been formed should end up in a rotating flattened configuration 
\citep[see, e.g.,][and references therein]{Larson2003}. Thus, it explains why the AB Aur envelope is 
flattened and suggests that the envelope slowly rotates, $V_{\rm rot} \la 0.1$~km\,s$^{-1}$, in addition to the 
infall. 

Furthermore, by fitting the line widths with various microturbulent velocities, we find that 
turbulence in the envelope should be low, $V_{\rm turb}\la0.2$~km\,s$^{-1}$, which is smaller than the 
sound speed, $c_s\approx0.4$~km\,s$^{-1}$. Therefore, these subsonic turbulent motions cannot retard 
the cloud collapse. Also, the AB Aur envelope cannot be supported against the gravitational contraction
by thermal pressure either since it is not spherically symmetric, and rather cold, $T\la35$~K. 
Finally, magnetic pressure can slow down the collapse, but only if the envelope matter is 
well coupled to the magnetic field. The computed ionization fraction for the best-fit AB Aur envelope is 
about $10^{-8}$-$10^{-6}$ and thus the latter requirement is fulfilled.

Therefore, we conclude that slow rotation and magnetic pressure play a major role in the dynamical evolution
of the AB Aur envelope by regulating the speed of the collapse. In turn, it defines the accretion rate 
onto the disk and thus its mass and consequently the lifetime of the {\em entire} AB Aur system, $t \la 25$~Myr. 
Also, these two dynamical factors are responsible for the flattened appearance of the envelope. Note that the presence 
of such asymmetric structures around protoplanetary disks has been predicted by modern theories of star formation 
\citep[e.g.,][]{Larson2003} and inferred from observations \citep[e.g.,][]{Hog2003}. In a more general sense, 
our findings support the idea of \citet{Armitageea2003}, who have argued that the observed disk lifetimes between 
1 and 10~Myr are mostly determined by the initial mass available for the accretion, but not by the mass of the central 
star(s).

The instantaneous infall of the envelope matter onto the rotationally supported disk should produce
an accretion shock interface, and likely {\em local} density enhancement(s) at the disk outer edge 
\citep{Velusamyea2002}. Then it is natural to ask whether these disk regions remain stable or not.
The gravitational instability in rotating disklike configurations is suppressed when the so-called Toomre
parameter exceeds about unity \citep[see discussion in][]{Larson2003}:
\begin{equation}
 Q(r)=\frac{c_s\Omega(r)}{\pi G \Sigma(r)},
\end{equation}
where $c_s$ is the sound speed, $\Omega(r)$ is the epicyclical velocity at the radius $r$, $G$ is gravitational 
constant, and $\Sigma(r)$ being the disk surface density.

Applying this equation to three considered disk models that fit the observations, namely, the uniform disk 
($p=0$), minimum-mass solar nebula ($p=-3/2$), and reference model ($p=-5/2$), we find that the corresponding
$Q$-factors at the disk outer boundary $r=400$~AU are 0.57, 22, and 300, respectively. It is likely that these 
values are smaller because of the local density enhancement at the disk outer edge due to the mass flow 
from the envelope. Thus, continuous mass feeding of the AB Aur disk from the surrounding envelope may make it 
gravitationally unstable. Consequently, the disk will develop prominent spiral arms trailing around through which 
the matter from the outer edge will be rapidly transported inward, to the disk inner regions \citep{Larson2003}. 

In Sections~\ref{disk_res} and \ref{env_den} we have claimed that presence of local high-density (clump-like) structures 
in the AB Aur system is a necessary requirement to explain the peculiar shape of the HCO$^+$(1-0) profiles in a region
of the PdBI map and the high ratio of the single-dish IRAM CS(5-4) and CS(2-1) spectra. This suggestion is supported
by the near-infrared {\it Subaru} observations of \citet{Fea04} that reveal the presence of a few 
spiral arms, a knotty structure, a dark lane, and underlying circular inhomogeneous structure in the AB Aur system at 
$r \la 600$~AU (see Fig.~3 therein). They have also found that these arms are trailing and concluded from 
similar arguments that we caught the gravitational instability at work in a protoplanetary disk. Note that 
spiral density waves can also be produced by tidal interactions between the disk matter and 
hidden low-mass (planetary) companion(s). However, Fukagawa et al. have ruled out such a possibility because otherwise a companion would have been detected. It is worth to mention here that recently \citet{Fromang2004} have simulated the evolution
of magnetized self-gravitating disks in 2D and 3D that show amazingly similar spiral structure as has been observed
by \citet{Fea04} (compare Figures 5 and 3 there, respectively).

The continuous replenishment of the AB Aur disk matter by ``fresh'' material from the envelope via inward 
accretion by the spiral arms provides a natural explanation for the observational fact that the bulk of dust grains 
in the disk are pristine and closely resemble the ISM dust particles \citep[see][]{sil,Meeusea01}. Actually, it can also 
be true for some other Herbig Ae/Be systems as well. Though this hypothesis should be carefully verified by future 
observations and modeling, it can explain why some Herbig systems do not show significant grain evolution after 
several million years of the evolution, while other do.

\section{Summary and conclusions}
\label{con}

We observed the AB Aur system at millimeter wavelengths and studied its chemical composition with the IRAM 30-m antenna and 
Plateau de Bure array during 2000-2002. Overall, nine different molecular species in a dozen rotational transitions
were detected at low resolutions ($10$-$30\arcsec$) using the IRAM telescope: CO, C$^{18}$O, CS, HCO$^+$, DCO$^+$, 
H$_2$CO, HCN, HNC, and SiO. From the measured negative intensity of the DCO$^+$(2-1) line, we found strong evidence that 
the AB Aur envelope extends up to at least $\sim 35\,000$~AU from the star. In contrast, with PdBI we detected only the 
HCO$^+$(1-0) emission from the AB Aur disk at the modest $\sim5\arcsec$ resolution. The symmetric ``butterfly'' (two-lobe)
appearance of the intensity-weighted velocity map is indicative of a rotating $\la10\arcsec$ ($\la1\,500$~AU) disk that is 
seen close to face-on with a positional angle of $\sim90\degr$. To account for these observational data, we used {\em for 
the first time} a coherent modeling of the disk and envelope physical structure, chemical evolution, and 
radiative transfer in molecular lines.

First, we modeled the AB Aur disk by the 2D flared passive disk model with vertical temperature gradient and Keplerian 
rotation, using the 2D continuum radiative transfer code and observational facts and theoretical constraints from the 
literature. To represent the AB Aur envelope, we adopted the infalling isothermal spherical cloud model with a central 
region shadowed by the disk and two wide cones transparent to the stellar radiation. Second, for both the disk and envelope
models, time-dependent abundances and column densities of observationally important molecules were calculated for 3~Myr of 
the evolution with the gas-grain UMIST\,95 chemical network supplied by dust surface reactions, reactions of deuterium fractionation, and CO and H$_2$ shielding. After that, the calculated abundances of HCO$^+$, CO, C$^{18}$O, and CS molecules were translated 
to the excitation temperatures of the CO(2-1), C$^{18}$O(2-1), HCO$^+$(1-0), HCO$^+$(3-2), CS(2-1), and CS(5-4) transitions
by mean of the 2D non-LTE line radiative transfer code. Finally, with the same code we synthesized the beam-convolved 
HCO$^+$(1-0) interferometric map and single-dish CO(2-1), C$^{18}$O(2-1), HCO$^+$(1-0), HCO$^+$(3-2), CS(2-1), and 
CS(5-4) lines and compared them {\em directly} with the observational data. Iterating this modeling scheme about 30 times
(each run took $\sim3$ days on a Pentium IV 2.4GHz PC), we constrained the parameters of the AB Aur system and their 
uncertainties by varying the model configurations in a robust step-by-step way.

Overall success of such an advanced and complicated theoretical approach to explain the observational data is surprising.
The best-fit disk model reproduces the intensities, widths, and profiles of the observed HCO$^+$(1-0) spectra on the 
entire interferometric map apart from one corner. There the line profiles show a fake absorption-emission shape that 
resembles the inverse P Cygni profile. We suggested that this can be {\em indirect} evidence for a {\em local} inhomogeneity 
(density enhancement) of the nearby envelope at $r\ga600$~AU. 

The constrained parameters of the AB Aur disk are the following. The AB Aur disk is in Keplerian rotation and inclined by 
$\iota=17^{+6}_{-3}\degr$, whereas its position angle is $\phi=80\pm30\degr$. The uncertainties of the derived inclination 
angle are mainly caused by the uncertainty of the adopted stellar mass and spread in the radial gradient of the disk 
surface density. The radius of the disk is $R_\mathrm{out}=400\pm200$~AU and its mass is 
$M_\mathrm{disk}=1.3\cdot10^{-2}\,M_{\sun}$ with a factor of $\sim7$ uncertainty. The uncertainty of the constrained mass 
is mainly caused by possible variations of the gas-to-dust mass ratio and the size of dust grains in this object.

The best-fit model of the AB Aur disk and inner shadowed part of the envelope (disk-in-envelope model) successfully 
reproduces the intensities, widths, and profiles of the single-dish CO(2-1), C$^{18}$O(2-1), HCO$^+$(1-0), HCO$^+$(3-2), 
and CS(2-1) spectra with the exception of the CS(5-4) data, which can be fitted only by the model with clumps that have a
characteristic density of about $5\cdot 10^7$~cm$^{-3}$ and homogeneously fill in $3\%$ of the full $10\arcsec$ 
IRAM beam in the CS(5-4) transition. We found that the large $\sim2$~km\,s$^{-1}$ width of the observed 
CO(2-1) emission cannot be explained by this model alone and is likely due to contamination by moving gas clouds along the 
line of sight to AB Aur. The best-fit envelope model has a mean temperature of about $35\pm14$~K, power-law density 
distribution $\rho \propto r^{-1.0\pm0.3}$ with the initial density of $9.4\cdot10^{-19}$~g\,cm$^{-3}$ at $400$~AU, 
mass of the shielded region within $400<r<2\,200$~AU, $M^\mathrm{sh}_\mathrm{env}=4\cdot10^{-3}\,M_{\sun}$, and total 
mass of about $6\cdot10^{-3}\,M_{\sun}$ (the latter three quantities are uncertain by a factor of $\sim7$). Here, 
the $\pm14$~K uncertainty of the average envelope temperature is due to the $\sim10$-15$\%$ calibration errors in the 
observed CO(2-1) flux and a $\sim15\%$ uncertainty in the distance to AB Aur.

The estimated parameters of the AB Aur disk and envelope are in reasonable agreement with other studies performed so far.
Furthermore, the applied step-by-step theoretical approach allowed to account for various observed interferometric and 
single-dish molecular spectra of AB Aur {\em simultaneously}. We conclude that a comprehensive theoretical modeling of the observed
interferometric maps obtained even with modest resolution offers an unique possibility to constrain the masses, sizes, 
orientation, and dynamical structure of young protoplanetary disks in an {\em independent} way. The same approach applied 
to single-dish spectra is capable in determining average temperature, density, and kinematical structures of the 
surrounding envelopes. 

Moreover, our best-fit model predicts that the $\sim 10^{-8}M_{\sun}$~yr$^{-1}$ mass accretion to be in the AB Aur 
system is regulated by steady contraction of the envelope that is only partly supported by rotation and magnetic field
acting against gravitation. It also gives a rough estimate of the timescale for dispersal of the entire system, $t\la 25$~Myr. 
We argue that the continuous mass supply from the infalling envelope onto rotating disk produces gravitational 
instabilities, resulting in a spiral disk structure, similar to that recently observed by \citet{Fea04}
and simulated with a 3D MHD code by \citet{Fromang2004}. We conclude that the continuous replenishment of the AB Aur disk 
matter by the material from the surrounding envelope provides a straightforward explanation to the observational fact that most 
of the disk grains are pristine and resemble ISM dust particles even after several Myr of the evolution, which can also
be true for other Herbig Ae/Be systems.
 
\begin{acknowledgements}
DS is supported by the {\it Deutsche Forschungsgemeinschaft},
DFG project ``Research Group Laboratory Astrophysics'' (He 1935/17-2). YP
was financially supported by the RFBR grant 04-02-16637. 
Authors are thankful to the anonymous referee for valuable comments and
suggestions. Also, we acknowledge the help by the staff members at IRAM during the observations.
Furthermore, we gratefully thank Anne Dutrey and Boris Shustov for fruitful discussions.
This research has made use of NASA's Astrophysics Data System.
\end{acknowledgements}


\clearpage

\clearpage

\clearpage

\clearpage

\clearpage

\clearpage

\clearpage

\clearpage

\clearpage

\clearpage

\clearpage

\clearpage

\clearpage

\clearpage
\begin{deluxetable}{llccccccccc}
\tabletypesize{\scriptsize}
\tablewidth{0pt}
\rotate
\tablecaption{Parameters of the detected single-dish emission lines\label{lines}}
\tablehead{
\colhead{Molecule} & \colhead{Transition} & \colhead{$\nu$} & \colhead{$T_{\rm mb}$} & \colhead{rms} &
\colhead{$\Delta V_\mathrm{obs}$} & \colhead{V$_{\rm lsr}$} & \colhead{$\int T_{\rm mb}$dv} &\colhead{$D_{\rm v}$} &
\colhead{$\eta_{\rm beam}$} & \colhead{Beamsize} \\
\colhead{} & \colhead{($J$=)} & \colhead{(MHz)} & \colhead{(K)} & \colhead{(K)} & \colhead{(km\,s$^{-1}$)} &
\colhead{(km\,s$^{-1}$)} & \colhead{(K\,km\,s$^{-1}$)} & \colhead{(km\,s$^{-1}$)} & \colhead{} &
\colhead{($\arcsec$)} \\
}
\startdata
CO & $2-1^{\rm a}$ & 230538.0 & 25.2 & 0.176 & 1.2 & 5.9 & 32.5 & 0.051 & 0.53 & 10.8 \\
C$^{18}$O& $2-1^{\rm a}$ & 219560.3 & 0.84 & 0.135 & 0.4 & 5.9 & 0.33 & 0.053 & 0.55 & 11.4 \\
CS & $2-1^{\rm a}$ & 97981.0 & 0.116 & 0.030 & 0.4 & 5.8 & 0.045 & 0.120 & 0.76 & 25.2 \\
& $5-4^{\rm a}$ & 244935.1 & 0.44 & 0.158 & 0.20 & 4.8 & 0.136 & 0.048 & 0.5 & 10.0 \\
   & $5-4$ & 244935.1 & 0.73 & 0.190 & 0.17 & 4.7 & 0.136 & 0.048 & 0.5 & 10.05 \\
HCO$^+$ & $1-0$ & 89188.5 & 0.54 & 0.029 & 0.54 & 5.9 & 0.310 & 0.066 & 0.78 & 29.0 \\
 & $2-1^{\rm b}$ & 178375.0 & 1.41 & 0.417 & 0.27 & 3.7 & 0.404 & 0.066 & 0.65 & 14.0 \\
 & $3-2^{\rm b}$ & 267557.6 & 1.16 & 0.269 & 0.26 & 4.8 & 0.315 & 0.044 & 0.45 & 9.3\\
DCO$^+$ & $2-1^{\rm a}$  & 144077.3 & 0.22 & 0.077 & 0.17 & 5.6 & 0.040 & 0.081 & 0.69 &  17.1 \\
 & $2-1^{\rm c}$ & 144077.3 & $-1.13$ & 0.033 & 0.17 & 6.2 & $-0.024$ & 0.081 & 0.69 & 17.1 \\
HCN & $1-0$, F$=0-1$ & 88633.9 & 0.117 & 0.025 & 0.65 & 9.9 & 0.080 & 0.132 & 0.78 & 28.2\\
 & $1-0$, F$=2-1$ & 88631.8 & 0.201 & 0.025 & 0.77 & 5.1 & 0.165 & 0.132 & 0.78 & 28.2\\
HNC & $1-0^{\rm a}$ & 90663.6 & 0.113 & 0.020 & 1.29 & 6.3 & 0.155 & 0.129 & 0.77 & 27.6\\
SiO & $2-1$, v$=0$ & 86847.0 & 0.091 & 0.027 & 0.19 & 7.0 & 0.019 & 0.067 & 0.78 & 29.0\\
H$_2$CO & ${\rm K_p,K_o}=2_{1,2}-1_{1,1}$ & 140839.5 & 0.234 & 0.089 & 0.25 & 6.0 & 0.063 & 0.083 & 0.70 & 17.5 \\
 & ${\rm K_p,K_o}=3_{1,2}-2_{1,1}$ & 225697.8 & 0.899 & 0.232 & 0.27 & 5.2 & 0.264 & 0.133 & 0.54 & 11.0\\
\enddata
\tablecomments{The Col.~(2) lists observed transitions at the rest frequencies from Col.~(3). The intensities,
noise, and FWHMs of detected lines are presented in Col.~(4)--(6), while in Col.~(7) the line velocity shifts
are specified. The corresponding integrated intensities and achieved spectral resolutions are listed in
Col.~(8) and Col.~(9). Finally, in Col.~(10)--(11) the beam efficiencies and sizes are given.}
\tablenotemark{a}\tablenotetext{a}{Frequency switching measurements,}
\tablenotemark{b}\tablenotetext{b}{Fit is made to the strongest line of a double-peak profile,}
\tablenotemark{c}\tablenotetext{c}{Signal is negative due to subtraction of the strong DCO$^+$(2-1) signal arising in
the outer cold envelope region from the weaker emission that originates in the inner warm (and thus less deuterated) part 
of the AB Aur envelope.}
\end{deluxetable}

\clearpage
\begin{deluxetable}{llll}
\tablewidth{0pt}
\tablecaption{Parameters of the central star\label{star_par}}
\tablehead{\colhead{Parameter} & \colhead{Symbol} & \colhead{Value} & \colhead{Reference}}
\startdata
Distance & $r_*$ & $144^{+23}_{-17}$~pc & 1 \\
Temperature & $T_\mathrm{eff}$ & $10\,000\pm500$K & 2 \\
Radius & $R_*$ & $2.5\pm0.2R_{\sun}$ & 3 \\
Mass & $M_*$ & $2.4\pm0.2M_{\sun}$ & 2 \\
UV flux at 100~AU & $G_*$ & $1.5\cdot10^5$ & 4 \\
\enddata
\tablerefs{(1) van den Ancker et al.~1997; (2)  van den Ancker et al.~1998; (3)  van den Ancker et al.~2000; 
(4) this work}
\end{deluxetable}

\clearpage
\begin{deluxetable}{llll}
\tablewidth{0pt}
\tablecaption{Parameters of the best-fit disk model\label{disk_par}}
\tablehead{\colhead{Parameter} & \colhead{Symbol} & \colhead{Value}}
\startdata
Inner radius & $R^\mathrm{disk}_\mathrm{in}$ & 0.7~AU\\
Outer radius & $R^\mathrm{disk}_\mathrm{out}$ & $400\pm200$~AU\\
Temperature & $T_\mathrm{disk}$ & 30-1\,500K\\
Surface density at 1~AU & $\Sigma_0$ & $7.7\cdot10^3$~g\,cm$^{-2}$ ($\pm$ factor of $\sim 7$)\\
Surface density profile & $p$ & $-2.5$\\
Mass & $M_\mathrm{disk}$ & $1.3\cdot10^{-2}M_{\sun}$ ($\pm$ factor of $\sim 7$)\\
Grain radius & $a_\mathrm{disk}$ & $0.3\mu$m\\
Gas-to-dust mass ratio & $m_\mathrm{gd}$ & 100\\
\enddata
\end{deluxetable}

\clearpage
\begin{deluxetable}{llll}
\tablewidth{0pt}
\tablecaption{Parameters of the best-fit envelope model\label{env_par}}
\tablehead{\colhead{Parameter} & \colhead{Symbol} & \colhead{Value}}
\startdata
Inner radius & $R^\mathrm{env}_\mathrm{in}$ & $400\pm200$~AU\\
Outer radius & $R^\mathrm{env}_\mathrm{out}$ & $2\,200$~AU$^{\rm \bf a}$\\
Shadowing angle & $\theta$ & $25\degr$\\
Temperature & $T_\mathrm{env}$ & $35\pm14$K \\
Density at 400~AU & $\rho_0$ & $9.4\cdot10^{-19}$~g\,cm$^{-3}$ ($\pm$ factor of $\sim 7$)\\
Density profile & $p$ & $-1.0\pm0.3$\\
Mass & $M^\mathrm{sh}_\mathrm{env}$ & $4\cdot10^{-3}M_{\sun}$ ($\pm$ factor of $\sim 7$)\\
Grain radius & $a_\mathrm{env}$ & 0.1~$\mu$m\\
Gas-to-dust mass ratio & $m_\mathrm{gd}$ & 100\\
\enddata
\tablenotemark{a}\tablenotetext{a}{This value is limited by the largest $29\arcsec$ IRAM beam size used in
our observations.}
\end{deluxetable}

\clearpage
\begin{deluxetable}{lll}
\tablewidth{0pt}
\tablecaption{Parameters of the 2D-LRT calculations\label{lrt_par}}
\tablehead{\colhead{Parameter} & \colhead{Symbol} & \colhead{Value}}
\startdata
Resolution of the numerical grid:  & $r, \theta$ & ...\\
1) Disk model & ... & $28\times55$ \\
2) Envelope model & ... & $28\times55$ \\
3) Disk-in-envelope model & ... & $56\times55$\\
Error in the level populations & ... & $\la5\%$ \\
System velocity & $V_\mathrm{lsr}$ & $5.85\pm0.1$~km\,s$^{-1}$ \\
Microturbulent velocity & $V_\mathrm{turb}$ & $0.2$~km\,s$^{-1}$ \\
Disk regular velocity & $V_\mathrm{disk}$ & $46.2\cdot(r/1\mathrm{AU})^{-0.5}$~km\,s$^{-1}$\\
Envelope regular velocity & $V_\mathrm{env}$ & $0.2\cdot(r/400\mathrm{AU})^{-1}$~km\,s$^{-1}$\\
Background temperature & $T_\mathrm{bgr}$ & 2.73~K \\
Spectral resolution & $D_\nu$ & 0.04 km\,s$^{-1}$ \\
Inclination angle & $\iota$ & $17^{+6}_{-3}\degr$ \\
Positional angle & $\phi$ & $80\pm30\degr$ \\
PdBI HPBW beam & ... & $5.87\arcsec$\\
IRAM HPBW beam & ... & $9.3\arcsec$-$29\arcsec^\mathrm{a}$\\
\enddata
\tablenotemark{a}\tablenotetext{a}{See Table~\ref{lines}, Col.~(11)}
\end{deluxetable}

\clearpage
\begin{deluxetable}{llll}
\tablewidth{0pt}
\tablecaption{Disk mass as a function of model parameters\label{disk_mass}}
\tablehead{
\colhead{Parameter} & \colhead{Symbol} & \colhead{Value}  & \colhead{Disk mass$^{\rm a}$}
}
\startdata
Gas-to-dust ratio       & $m_\mathrm{gd}$                & 10                 & 0.2 \\
Gas-to-dust ratio       & $m_\mathrm{gd}$                & 1000               & 5.0 \\ 
Grain radius            & $a_\mathrm{disk}$              & $1\mu$m            & 5.0 \\
HCO$^+$ intensities     & $T_\mathrm{mb}$                & $\la0.14$~K        & 0.3 \\
HCO$^+$ intensities     & $T_\mathrm{mb}$                & $\la0.3$~K         & 2.1 \\
HCO$^+$ abundances      & ...                            & 2 of standard      & 0.21 \\
HCO$^+$ abundances      & ...                            & 0.5 of standard    & 1.8 \\
Surface density profile & $p$                            & 0                  & 0.7 \\
Outer radius            & $R^\mathrm{disk}_\mathrm{out}$ & 400~AU             & 1.5 \\
\enddata
\tablenotemark{a}\tablenotetext{a}{Disk masses are given in units of the best-fit mass 
$M_\mathrm{disk}=1.3\cdot10^{-2}M_{\sun}$.}
\end{deluxetable}

\clearpage
\begin{deluxetable}{lllll}
\tablewidth{0pt}
\tablecaption{Comparison of the AB Aur envelope models\label{env_com}}
\tablehead{
  \colhead{Study} & \colhead{Initial density at} & \colhead{Density} & \colhead{Mass$^{\rm a}$} & \colhead{Total mass}\\
  \colhead{} & \colhead{400~AU, g~cm$^{-3}$} & \colhead{profile} & \colhead{$M_{\sun}$} & \colhead{$M_{\sun}$}
}
\startdata
1 & $2.3\cdot10^{-20}$ &  0   & $5\cdot10^{-4}$   & $1.6\cdot10^{-3}$ \\
2 & $9.4\cdot10^{-19}$ & -1   & $4\cdot10^{-3}$   & $6\cdot10^{-3}$ \\
3 & $1.3\cdot10^{-18}$ & -1   & $6.8\cdot10^{-3}$ & $2.5\cdot10^{-2}$ \\
4 & $1.6\cdot10^{-18}$ & -1.4 & $5.6\cdot10^{-3}$ & $2\cdot10^{-2}$ \\
3 & $3.7\cdot10^{-17}$ & -2   & $5.4\cdot10^{-2}$ & $2\cdot10^{-1}$ \\
\enddata
\tablenotemark{a}\tablenotetext{a}{Mass of the shadowed ($\approx28\%$ by volume) part of the envelope between
400~AU and $2\,200$~AU.}
\tablerefs{(1) Miroshnichenko et al. 1999; (2) This work; (3) Fuente et al. 2002; (4) Elia et al. 2004}
\end{deluxetable}

\end{document}